\documentclass[aps,prb,preprint,notitlepage]{revtex4-2}
\usepackage{amsmath}
\usepackage{graphicx}
\usepackage{amssymb}

\usepackage{gt17}
\usepackage{color} 
\usepackage{graphicx}
\usepackage{bm}
\usepackage{amsmath}
\begin{document}
\title{Hydrodynamic theory of vorticity-induced  spin transport
}
\author{Gen Tatara} 
\affiliation{RIKEN Center for Emergent Matter Science (CEMS)
and RIKEN Cluster for Pioneering Research (CPR), 
2-1 Hirosawa, Wako, Saitama, 351-0198 Japan}
\date{\today}
\begin{abstract}
Electron spin transport in a disordered metal is theoretically studied from the hydrodynamic viewpoint focusing on the role of electron vorticity.
The spin-resolved momentum flux density of electrons is calculated taking account of the spin-orbit interaction and uniform magnetization, and  the expression for the spin motive force is obtained as the linear response to a driving electric field.
It is shown that the spin-resolved momentum flux density and motive force are characterized by troidal moments expressed as vector products of the applied external electric field and  the spin polarization and/or magnetization. 
The spin-vorticity and magnetization-vorticity couplings studied recently are shown to arise from the toridal moments contribution to the momentum flux density. 
Spin motive force turns out to have a nonconservative contribution besides the conventional conservative one due to the spin-vorticity coupling.   
Spin accumulation induced by an electric field is calculated to demonstrate the direct relation between vorticity and induced spin, 
and the spin Hall effect is interpreted as due to the spin-vorticity coupling. 
The spin-vorticity coupling is shown to give rise to a vorticity-induced torque and a spin relaxation.  
The vorticity-induced torque is a linear effect of the spin-orbit interaction and is expected to be larger than the second-order torques such as nonadiabatic ($\beta$) current-induced torque due to magnetization structure.  
The intrinsic inverse spin Hall effect is argued to correspond to the antisymmetric components of the momentum flux density in the hydrodynamic context. 
\end{abstract}

\maketitle

\newcommand{\chiv}{{\bm \chi}}
\newcommand{\fv}{{\bm f}}
\newcommand{\omegav}{{\bm \omega}}
\newcommand{\omegavE}{\omegav}
\newcommand{\omegaE}{\omega}
\newcommand{\sigmae}{\sigma_{\rm e}}
\newcommand{\Tv}{{\bm T}}
\newcommand{\Mvhat}{\hat{\Mv}}
\newcommand{\svhat}{\hat{\sv}}

\section{Introduction}

Electron transport in magnetic metals has been studied in various systems. 
Technologically most successful effect would be the giant magnetoresistance (GMR)  effect discovered in magnetic layers \cite{Baibich88}. 
GMR effects in multilayers were theoretically described based on a two-current electron model derived from the Boltzmann equation for the conduction electrons with spin up and down \cite{Valet93}. 
Effects of magnetic structures have been argued in the context of magnetoresistance  and spin-transfer torque \cite{Berger78,Berger86,TKS_PR08}.
In the previous works on the spin transport, effects of nonuniform magnetization were focused on, while current density was treated as uniform. 
In reality, however, the current is generally inhomogeneous at edges, resulting in local vorticity of electron flow.
Vorticity of current density $\jv$, $\nabla\times \jv$, carries angular momentum and may couple to the magnetic field and spin, affecting local spin transport. 
In fact, electron's vorticity was shown theoretically to couple to electron spin via an quantum relativistic effect \cite{Matsuo11}, and various  phenomena induced by the spin-vorticity coupling have been discussed recently \cite{MatsuoHydro17,Doornenbal19,Fujimoto21}. 

Vorticity arises generally near surfaces and interfaces where longitudinal flow is suppressed, and vorticity-induced effects are expected to dominate various transport properties in thin films and mesoscopic systems. 
For discussing vorticity effects, hydrodynamic equations are useful to identify the forces due to inhomogeneity of the flow. 
Electron transport in metals with disorder is governed by relaxation force instead of viscosity force in conventional fluids.  
Coarse grained behaviors  of such disordered metals are described in terms of an ohmic fluid \cite{Gurzhi63}.
Hydrodynamic description of ohmic electron fluid was employed to describe angular momentum generation in chiral electron systems and anomalous Hall effect \cite{Funaki21,FunakiAH21}.
Hydrodynamic equations are conventionally represented in terms of current density $\jv$.  
In ohmic fluids, electric current density is locally related to a driving electric field  $\Ev$ as $\jv=\sigmae \Ev$, where $\sigmae$ is a conductivity tensor. 
Using the local relation, the hydrodynamic equations can thus be represented in terms of the driving field. 
In the driving-field representation, hydrodynamic coefficients are directly related to microscopic response functions to the applied field, and are systematically calculable by use of a microscopic linear response theory \cite{Conti99,Funaki21}. 

In this paper, we study spin transport effects from the hydrodynamic viewpoints focusing on the effects of vorticity. 
We calculate hydrodynamic coefficients as a linear response to the applied field  taking account of its inhomogenuity to the lowest order of spatial derivative.

\subsection{Spin-vorticity coupling}
As was argued in Refs. \cite{Matsuo11,MatsuoHydro17}, spin-vorticity coupling is an interaction proportional to 
$\sv\cdot(\nabla\times\jv)$, the scalar product of spin density $\sv$ and vorticity. 
The coupling induces a force on electron spin proportional to the gradient of vorticity and generates spin current, resulting in 
a spin accumulation at edges where spin current terminates. 
The vorticity-induced spin current can be detected by use of the inverse spin Hall effect \cite{MatsuoHydro17}. 

It was demonstrated recently \cite{TataraSH18} that this spin-vorticity coupling represents the spin Hall effect \cite{Dyakonov71,Hirsch99}. 
In fact, the spin Hall effect was shown to be represented by a single equation  
\begin{align}
\sv=\lambda_{\rm sh}\omegavE 
\label{sh}
\end{align}
where $\omegavE\equiv\nabla\times\Ev$ is a vorticity of the applied field and $\lambda_{\rm sh}$ is a constant arising from the spin-orbit interaction \cite{TataraSH18}. 
This relation is equivalent to the spin-vorticity coupling at the lowest order in the spin-orbit interaction due to the local relation between $\jv$ and $\Ev$.  
In the disordered case, spin density propagates by diffusion and the formula (\ref{sh}) is modified to include a diffusion propagator (See Eq. (\ref{snonlocal})). 
The expression is then equivalent to the conventional representation of spin accumulation in terms of spin diffusion equation. 

The spin-vorticity coupling potential induces a force proportional to its gradient, $\nabla(\sv\cdot\omegav)$. 
From the symmetry, another form of the force $(\sv\cdot\nabla)\omegav\equiv \fv_{\rm nc}$ is allowed. 
Noting $\nabla\cdot\omegav=0$, $\fv_{\rm nc}=\nabla\times(\omegav\times\sv)$ is a nonconservative force, which is allowed if there are  viscosity or friction. 
We shall demonstrate that such a nonconservative  force indeed exists in the present electron spin fluid. 
In the context of magnetization-vorticity coupling, the same form of a nonconservative force with $\sv$ replaced by magnetization was identified in Ref. \cite{FunakiAH21}. 

The spin-vorticity coupling indicates that there are current-induced torques arising from vorticity of current in ferromagnets. 
The torque arises from the inhomogeneity of current density instead of inhomogeneity of magnetization for conventional current-induced torques. 
The effect would be localized near surfaces and interfaces and is expected to be enhanced by introducing artificial roughness or inhomogeneous structures. 
We shall demonstrate that the vorticity-induced torque arises theoretically at the linear order of the spin-orbit interaction, while conventional current-induced nonadiabatic torque ($\beta$ torque) is the second-order effect \cite{TKS_PR08}. 
The vorticity-induced torque is thus expected to have larger effects than conventional spin-orbit driven torques. 
The vorticity-induced torque has a vanishing bulk component, while  it acts in thin films as an alternating torque, resulting in a spin relaxation.

\section{Anomalous Hall fluid \label{SECah}}
We first argue electron transport in a ferromagnetic metal with a uniform magnetization, i.e., the anomalous Hall system,  studied from the hydrodynamic viewpoint in Ref. \cite{FunakiAH21}. 
In that paper, the electron momentum flux density was calculated diagrammatically in the presence of a uniform magnetization at the linear response to the applied electric field treated as non uniform. 
It was shown that there is a contribution to the viscosity coefficient arising from a coupling between the magnetization and vorticity of electron velocity, i.e., the magnetization-vorticity coupling. In terms of applied field $\Ev$, the coupling potential is $-\tilde{\zeta}_{M}^{\rm f}\Mv\cdot\omegavE$, where $\Mv$ is the magnetization vector and $\omegavE\equiv \nabla\times\Ev$ is a vorticity  of the applied field and $\tilde{\zeta}_{M}^{\rm f}$ is a coefficient.
The force density due to the coupling is  
\begin{align}
\fv_{M\omega}=\tilde{\zeta}_{M}^{\rm f}\nabla(\Mv\cdot\omegavE) \label{fMomega}
\end{align}
Such a coupling has been argued from the phenomenological  ground \cite{Snider67} and derived microscopically \cite{FunakiAH21}.
It is regarded as a limit of spin-vorticity coupling where spin has a finite expectation value.
Electron fluid in a uniform external magnetic field was studied in Ref. \cite{Scaffidi17}.

Phenomenologically, the magnetization-vorticity coupling is understood simply taking account of the anomalous Hall effect in the conventional nonmagnetic fluid. 
Fluid dynamics is described by the momentum flux density, $\pi_{ij}$, which is an expectation value  $\average{\hat{p}_i \hat{v}_j}$ of momentum  and velocity operators,  $\hat{p}$ and $\hat{v}$, respectively ($i,j$ are spatial directions). 
Its divergence is the force density for the fluid, $f_i=-\nabla_j\pi_{ij}$. 
In nonmagnetic fluids with high symmetry,  $\pi_{ij}$ is written in terms of current density $\jv$ as 
$\pi_{ij}={\zeta}_0\delta_{ij}(\nabla\cdot\jv)+{\eta}_0[\nabla_ij_j+\nabla_j j_i]$, where ${\zeta}_0$  and ${\eta}_0$ are viscosity constants \cite{LandauLifshitz-FluidMechanics}.
The momentum flux density may have antisymmetric component, $\pi_{ij}^{\rm a}$, which is written generally as 
$\pi_{ij}^{\rm a}=\epsilon_{ijk}a_k$, where $\av$ is a vector invariant by the parity inversion ($\rv\ra-\rv$). 
Without broken symmetry, vorticity $\nabla\times\jv$ is allowed as the vector $\av$, resulting in an antisymmetric component 
$\pi_{ij}^{\rm a}=\frac{\xi_0}{2}(\nabla_ij_j-\nabla_j j_i)$ with a constant  $\xi_0$ \cite{Groot11}. 
The electron current density is proportional to the applied field as $\jv=\sigma_{\rm e}\Ev$, where $\sigma_{\rm e}$ is the diagonal conductivity  in the case of high symmetry. The momentum flux density in this case is therefore    
\begin{align}
\pi_{ij}=\zeta\delta_{ij}(\nabla\cdot\Ev)+ {\eta}[\nabla_i E_j+\nabla_j E_i] +\epsilon_{ijk}a_k
\label{piijnormal}
\end{align}
 with $\zeta\equiv {\zeta}_0\sigma_{\rm e}$ and  $\eta\equiv {\eta}_0\sigma_{\rm e}$ and $\av={\xi}\omegavE$ 
 ($\xi\equiv \xi_0\sigma_{\rm e}$). 
 
In the presence of magnetization, the anomalous Hall effect tends to distort electron motion towards a perpendicular direction, i.e., $\jv$ is modified to be $\jv+\alpha_{M}(\hat{\Mv}\times\jv)$, where $\alpha_{M}$ is a constant and $\hat{\Mv}\equiv \Mv/|M|$.
This corresponds to emergence of a perpendicular driving field  $\alpha_{M}(\hat{\Mv}\times\Ev)$.  
The anomalous Hall effect thus induces  new components of the momentum flux density, written in terms of 
a vector called a troidal moment
\begin{align}
\hat{\Mv}\times\Ev\equiv \Tv_{\Mvhat},
\end{align}
as
\begin{align}
\pi^M_{ij}=\zeta_{M}\delta_{ij}(\nabla\cdot\Tv_{\Mvhat})+ {\eta_{M}}[\nabla_i T_{{\Mvhat},j}+\nabla_j T_{{\Mvhat},i}] 
  +\epsilon_{ijk}a_{M,k}.
\label{piahT}
\end{align}
A possible vector $\av_{M}$ in the anomalous Hall fluid is 
\begin{align}
 \av_{M} &=  \xi_{M}^{\omega} (\Mvhat\times \omegavE) + \xi_{M}^{{\rm v}} \Mvhat(\nabla\cdot\Ev) 
      + \xi_{M}^{T} (\nabla\times\Tv_{\Mvhat})
  \label{avah}
\end{align}
where $\xi_{M}^{\omega}$, $\xi_{M}^{{\rm v}} $ and  $\xi_{M}^{T}$ are coefficients, representing 
the anomalous Hall component of vorticity, volume change and troidal moment, respectively.  
 In the microscopic calculation in the ohmic regime in Ref. \cite{FunakiAH21}, 
 $\xi_{M}^{\omega}$ and $\xi_{M}^{{\rm v}}$ arise from the side-jump process, while $ \xi_{M}^{T}=0$. 
Using $\nabla\times\Tv_{\Mvhat}=\Mvhat(\nabla\cdot\Ev)-(\Mvhat\cdot\nabla)\Ev$ for the present case of uniform $\Mvhat$, different representations of $ \av_{M}$ are possible.
In metals, $\nabla\cdot\Ev=0$, and  $\nabla\times\Tv_{\Mvhat}=-(\Mvhat\cdot\nabla)\Ev$, resulting in 
\begin{align}
 \av_{M} &=  \xi_{M}^{\omega} \nabla (\Mvhat\cdot \Ev) 
      + \tilde{\xi}_{M}^{T} (\nabla\times\Tv_{\Mvhat})
  \label{avah2}
\end{align}
where $\tilde{\xi}_{M}^{T}  \equiv {\xi}_{M}^{T} + {\xi}_{M}^{\omega}  $.
The first term $\xi_{M}^{\omega}$ has no physical effect because its force density vanishes ($\nabla\times\nabla=0$). 

The  force density calculated from Eq. (\ref{piahT}) is 
\begin{align}
f_{M,i} & \equiv  -\nabla_j \pi^M_{ij}
\nnr 
& = -\zeta_M^{f}\nabla_i(\nabla\cdot\Tv_{\Mvhat}) - \eta_M^f \nabla^2 T_{{\Mvhat},i}  
+\xi_{M}^{\omega}(\Mvhat\cdot\nabla){\omegavE}_i
  +\xi_{M}^{\rm v} (\Mvhat\times \nabla)_i(\nabla\cdot\Ev)
\label{fah}
\end{align}
where $\zeta_M^{f}\equiv\zeta_{M}+\eta_{M}+ \xi_{M}^{T}$, 
$ \eta_M^f \equiv {\eta_{M}} - \xi_{M}^{T}$. 
The results were presented in Ref. \cite{FunakiAH21} without using $\Tv_{\Mvhat}$. 
In the case of uniform magnetization, 
\begin{align}
\nabla\cdot\Tv_{\Mvhat} 
=-(\Mvhat\cdot\omegavE) 
\end{align}
 and thus the first term of Eq. (\ref{fah}) ($\nabla_i(\nabla\cdot\Tv_{\Mvhat}))$ represents the magnetization-vorticity coupling (Eq. (\ref{fMomega})). 
In metals, using incompressibility $\nabla\cdot\Ev=0$, 
$\nabla^2 \Tv_{{\Mvhat}}=-\nabla(\Mvhat\cdot\omegavE)+(\Mvhat\cdot\nabla)\omegavE$ and  the total force density
is written in terms of vorticity as 
\begin{align}
\fv_{M} 
&  =  \tilde{\zeta}_M^{f}\nabla (\hat{\Mv}\cdot\omegavE)
+\tilde{\xi}_{M}^{\omega}(\hat{\Mv}\cdot\nabla){\omegavE}
\label{fahall}
\end{align}
where $\tilde{\zeta}_M^{f}\equiv \zeta_M^{f}+  \eta_M^f $ and 
$\tilde{\xi}_{M}^{\omega} \equiv \xi_{M}^{\omega} -  \eta_M^f$.
The first term on the right-hand side of Eq. (\ref{fahall}) is a conservative force arising from the magnetization-vorticity coupling, while the second term is a nonconservative force, $\tilde{\xi}_{M}^{\omega}[\nabla\times({\omegavE}\times\hat{\Mv})]$. 
This result, Eq. (\ref{fahall}), was presented by a microscopic calculation in Ref. \cite{FunakiAH21}.
The two contributions, conservative and nonconservative, to the vorticity-induced motive force leads to an anisotropic motive force with respect to the direction of $\Mvhat$ 
(see Sec. \ref{SEC:phenomenological}).

\section{Spin-resolved momentum flux density and spin motive force}
In this section, we consider the spin transport, i.e., the spin-resolved hydrodynamic equation for electrons.
Including a spin direction of electron, the hydrodynamic coefficients have more possibilities \cite{Snider67}. 
The momentum flux density is defined spin-dependent as $ \pi_{ij}^{s,\alpha} \equiv \average{\hat{p}_i \hat{v}_j \sigma_\alpha}$, where $\sigma_\alpha$ is the Pauli  matrix ($\alpha$ denotes spin direction). 
The time-derivative of the spin-resolved momentum density 
\begin{align}
p_i^{\alpha}\equiv \average{\hat{p}_i\sigma_\alpha} 
\end{align}
 is represented in terms of $ \pi_{ij}^{s,\alpha}$ as 
\begin{align}
\dot{p}_i^{\alpha}=-\nabla_j \pi_{ij}^{s,\alpha}. 
\label{spindepequ}
\end{align}
We introduce a unit vector $\svhat$ to represent the spin direction as 
\begin{align}
\bm{\pi}^s_{ij} \equiv (\pi_{ij}^{s,x},\pi_{ij}^{s,y},\pi_{ij}^{s,z}) 
\equiv \svhat \overline{\pi}_{ij}^{s}
\label{svhatdef}
\end{align}
 where $\overline{\pi}_{ij}^{s}\equiv ((\pi_{ij}^{s,x})^2+(\pi_{ij}^{s,y})^2+(\pi_{ij}^{s,z})^2)^{1/2}$.

Effect of spin (called the internal angular momentum in Ref. \cite{Snider67})  on the electron fluid was theoretically discussed on a symmetry basis in Ref. \cite{Snider67}. 
The possible contributions of the total (spin-neutral) momentum flux density when spin is present were argued. 

Here we represent the momentum flux density as spin-polarized introducing a spin direction $\svhat$.
The spatial derivative of the spin-resolved momentum flux density represents a force acting on spin polarization, i.e., the spin motive force.
The spin motive force is a convenient quantity to discuss  spin current generation.

One needs, however, to understand that spin-resolved quantities $p_i^{\alpha}$ and $\bm{\pi}^s_{ij} $ are not directly measurable and are not physical observables. 
In other words, the flux density $\bm{\pi}^s_{ij} $ is not conserved and cannot be defined uniquely, as is in the case of spin current. 
Explicitly, ${\pi}^{s,\alpha}_{ij}$ is essentially an expectation value of  $ \average{\hat{p}_i \hat{v}_j \sigma_\alpha}$, i.e., the velocity ($\hat{v}$) of momentum and spin (See Eqs. (\ref{pizerodef})(\ref{pideltadef})), but its expression depends on the ordering of the three operators $\hat{p}_i$, $\hat{v}_j$ and $\sigma_\alpha$ and is thus not unique. 
Like in the case of spin Hall effect, predictions for experiments need to be provided in terms of physical spin accumulation or charge current by solving the hydrodynamic equation  
Eq. (\ref{spindepequ}).
Here in this paper, we discuss spin-resolved momentum flux density for understanding roles of vorticity on the spin transport and 
leave explicit calculations of hydrodynamic equations as a future work.
Spin density calculation is carried out independently from the  momentum flux density analysis in Sec. \ref{SECs}

\subsection{Phenomenological argument \label{SEC:phenomenological}}
\begin{figure}
 \includegraphics[width=0.3\hsize]{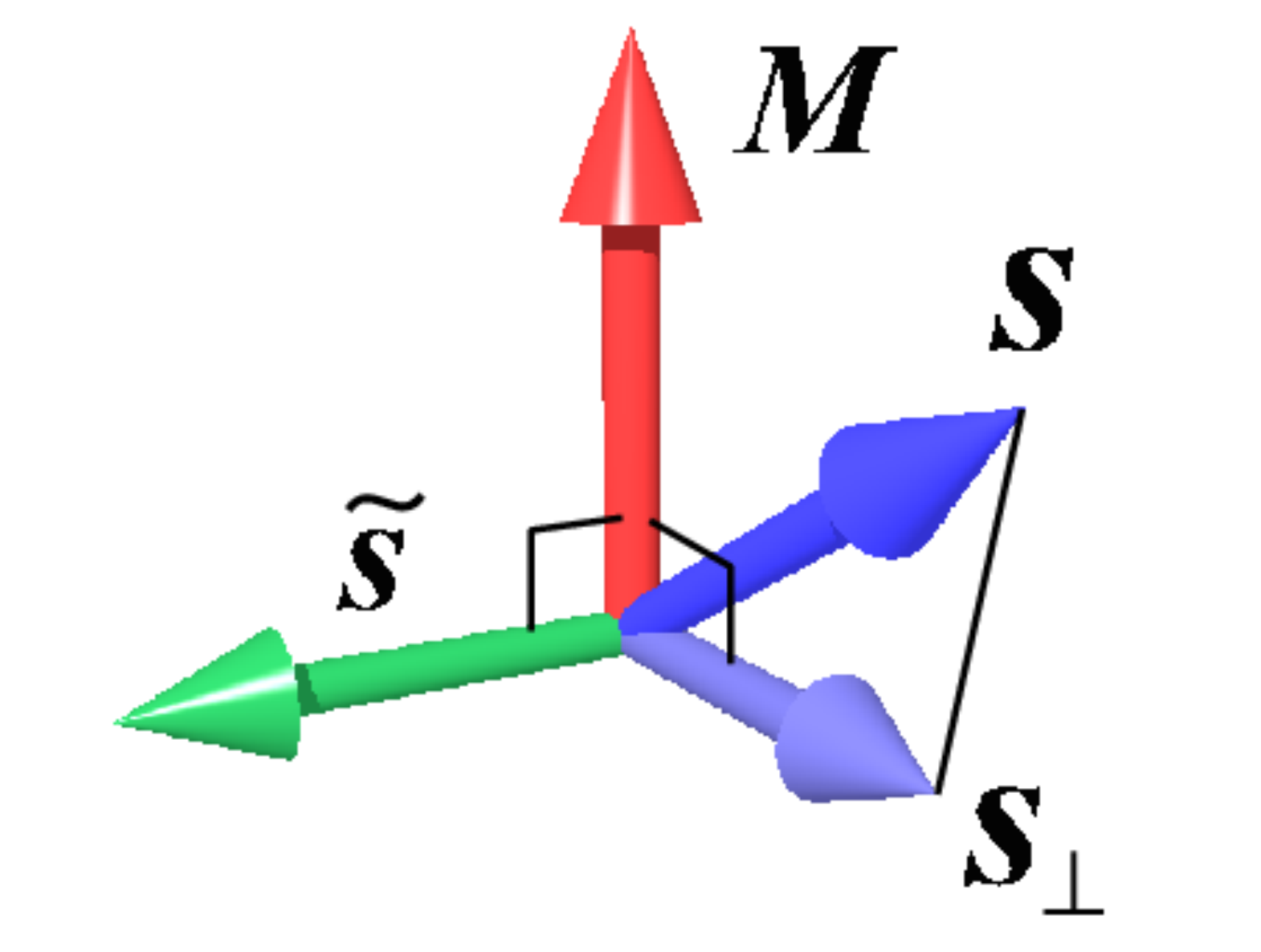}
 \caption{Parameterization of spin vectors, $\sv_\perp\equiv{\sv}-\hat{\Mv}(\hat{\Mv}\cdot{\sv})$ and 
 $\tilde{\sv}\equiv\sv \times\Mvhat$.
\label{FIGsdef}}
\end{figure}

Let us first proceed phenomenologically. 
We first focus on the symmetric (s) component,  $ \bm{\pi}_{ij}^{s{\rm (s)}} $.
Based on the expression for the anomalous Hall system \cite{FunakiAH21} (Eq. (\ref{piahT})), we expect an adiabatic component having a spin polarization $\svhat$ along the magnetization  as  ($\eta_s^{\parallel}$ and $\zeta_{s}^{\parallel}$  are coefficients) 
\begin{align}
 \overline{\pi}_{ij}^{s{\rm (s)}{\parallel}} 
   &= (\svhat\cdot\hat{\Mv}) \lt( \zeta_s^{\parallel} (\hat{\Mv}\cdot\omegavE)\delta_{ij} 
   + \eta_s^{\parallel} (\nabla_i T_{{\Mvhat},j} +\nabla_j T_{{\Mvhat},i}) \rt)
\end{align}
As for the  nonadiabatic contribution due to perpendicular spin polarization, we introduce
\begin{align}
  {\svhat}_\perp \equiv {\svhat}-\hat{\Mv}(\hat{\Mv}\cdot{\svhat})
\end{align}
whose contribution is 
\begin{align}
 \overline{\pi}_{ij}^{s{\rm (s)}{\perp}} 
   &=  \zeta_s^{\perp}(\hat{\sv}_\perp \cdot\omegavE)\delta_{ij} 
   + \eta_s^{\perp}(\nabla_i T_{\hat{\sv}_\perp,j} + \nabla_j T_{\hat{\sv}_\perp,i}) 
\end{align}
where 
\begin{align}
 \Tv_{{\hat{\sv}_\perp}} \equiv \hat{\sv}_\perp \times \Ev
\end{align}
is a troidal moment due to the perpendicular spin polarization. 
Another spin polarization perpendicular to $\hat{\Mv}$ is 
\begin{align}
\tilde{{\sv}}\equiv \hat{\sv}\times \hat{\Mv}. 
\end{align}
It is a spin polarization induced by an anomalous Hall effect in the direction perpendicular to $\Mv$, and its 
 contribution would be 
\begin{align}
 \overline{\pi}_{ij}^{s{\rm (s)}{\perp'}} 
   &=  \zeta_s^{\perp'} (\tilde{{\sv}}\cdot\omegavE)\delta_{ij} 
   + \eta_s^{\perp'} (\nabla_i T_{\tilde{{\sv}},j} + \nabla_j T_{\tilde{{\sv}},i} ) 
\end{align}
where 
\begin{align}
 \Tv_{\tilde{{\sv}}} \equiv \tilde{{\sv}} \times \Ev
\end{align}
and $\zeta_s^{\perp'}$ and $\eta_s^{\perp'}$ are constants. 

Antisymmetric contributions with spin direction $\alpha$, 
$\pi_{ij}^{s,\alpha} \equiv \epsilon_{ijk}a_k^{s,\alpha}$,  
 are similarly argued.
We consider the case of $\nabla\cdot\Ev=0$. 
As we saw in Eq. (\ref{avah2}), only the troidal moment remains in this case. 
Besides the  adiabatic component proportional to Eq. (\ref{avah2}), we therefore have  
(with coefficients $\xi_s^{T_{\hat{\sv}}}$ and $\xi_s^{T_{\tilde{\sv}}} $)
\begin{align}
 a_k^{s,\alpha} &=  
   \xi_s^{T_{\hat{\sv}}} (\nabla\times \Tv_{\hat{\sv}})
  + \xi_s^{T_{\tilde{\sv}}} (\nabla\times \Tv_{\tilde{\sv}})
  \label{avah}
\end{align}
(Here coefficients are defined using  $\svhat$ instead of $\svhat_\perp$, i.e., neglecting the adiabatic component.)

Due to Eq. (\ref{spindepequ}), a spin-resolved force (a spin motive force)  acting on the electron spin polarized along direction $\alpha$ is
\begin{align}
 f_{s,i}^{\alpha} & \equiv -\nabla_j  \pi_{ij}^{s,\alpha} 
 \equiv \svhat \overline{f}_{s,i}
\end{align}
As was argued in Sec. \ref{SECah} for the case of anomalous Hall fluid, the motive force when $\nabla\cdot\Ev=0$ is written in terms of vorticity. 
The adiabatic contribution parallel to $\Mvhat$ is essentially the same as the anomalous Hall case (Eq. (\ref{fahall})), i.e.,
\begin{align}
\overline{\fv}_s^{{\parallel}}
&  = (\hat{\sv}\cdot\Mvhat)[\tilde{\zeta}^{f,{\parallel}}_{s}\nabla (\hat{\Mv}\cdot\omegavE)
+\tilde{\xi}_s^{\omega,{\parallel}}(\hat{\Mv}\cdot\nabla){\omegavE} ]
\label{fsad}
\end{align}
with constants $\tilde{\zeta}^{f,{\parallel}}_{s}$ and $\tilde{\xi}_s^{\omega,{\parallel}}$, 
while other contributions lead to motive force depending on the spin polarization as 
\begin{align}
\overline{\fv}_{s}^{{\perp}}
&  =  \tilde{\zeta}^{f,\perp}_{s}\nabla (\hat{\sv}\cdot\omegavE)
+\tilde{\xi}_s^{\omega,\perp}(\hat{\sv}\cdot\nabla){\omegavE} 
+\tilde{\zeta}^{f,\perp'}_{s}\nabla (\tilde{\sv}\cdot\omegavE)
+\tilde{\xi}_s^{\omega,\perp'}(\tilde{\sv}\cdot\nabla){\omegavE} 
\label{fsperp}
\end{align}
The contributions represented by coefficients 
$\tilde{\zeta}^{f,\perp}_{s}$ and $\tilde{\zeta}^{f,\perp'}_{s}$ are conservative forces due to coupling potential to vorticity, while
 $\tilde{\xi}_s^{\omega,\perp}$ and $\tilde{\xi}_s^{\omega,\perp'}$ represent nonconservative forces. 
Interestingly, conservative forces induce  spin polarization ($\svhat$ or $\tilde{\sv}$) along $\omegavE$ and  force along the gradient of $\omegavE$, while the noncoservative  ones induce force and spin in the direction of $\omegavE$ and the gradient of $\omegavE$, respectively.

\begin{figure}
 \includegraphics[width=0.5\hsize]{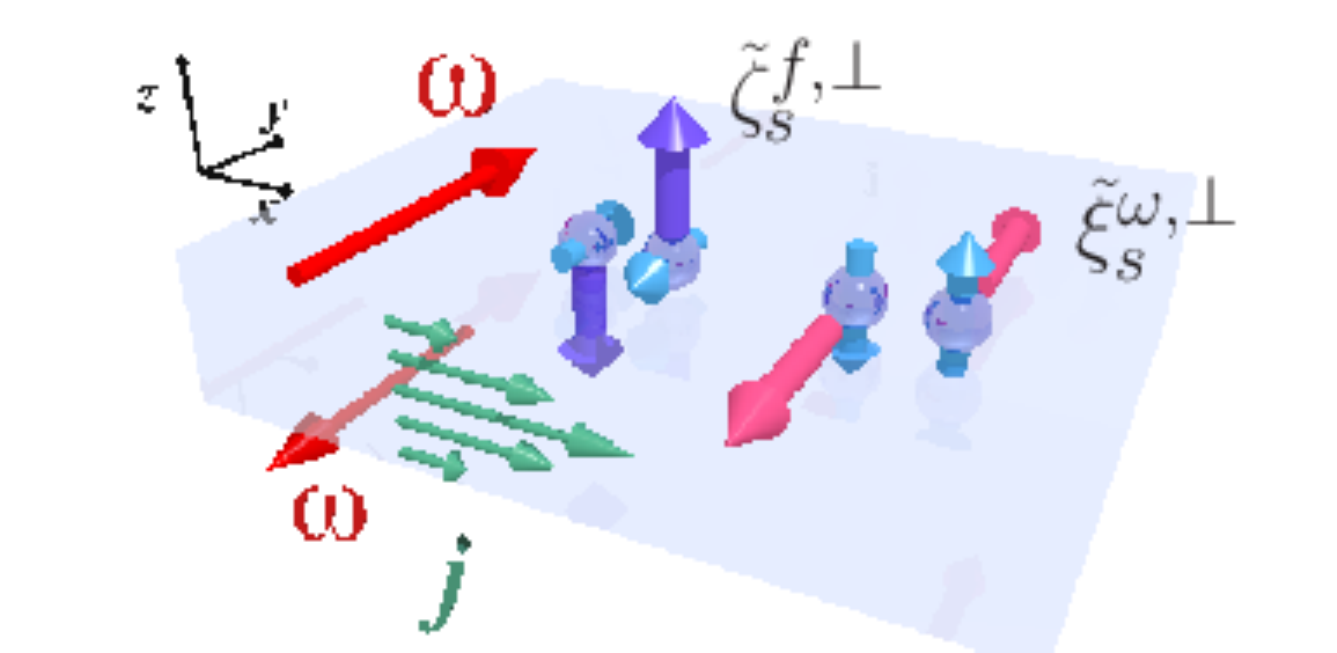}
 \caption{
A schematic picture showing spin motive forces of Eq. (\ref{fsperp}) in a thin film in the $xy$-plane with applied current along the $x$-direction. 
The contributions $\tilde{\zeta}^{f,\perp}_{s}$ and $\tilde{\xi}_s^{\omega,\perp}$  connecting vorticity and $\hat{\sv}$  are depicted. 
At the boundaries, the flow vanishes, meaning that total effective field $\Ev$ including the applied field and friction force from the boundary vanishes due to proportionality of $\jv$ and $\Ev$. 
The derivative $\partial_z E_x$ is therefore finite, namely, vorticity $\omega$ emerges near the boundary along the $y$ direction.  The vorticity has opposite signs on the upper and lower plane, resulting in $\partial_z \omega_y$. 
The contribution of conventional conservative spin-vorticity coupling, the term $ \tilde{\zeta}^{f,\perp}_{s}$, thus induces a spin polarization (short arrows with sphere) along the $y$ direction, and motive force along the $z$ axis (blue arrows).
The nonconservative contribution, $ \tilde{\xi}^{\omega,\perp}_{s}$, in contrast induces a spin polarization along the $z$ direction, and motive force along the $y$ axis (magenta arrows).
\label{FIGspinvorticity}}
\end{figure}

Let us consider electron fluid in a thin film as in Fig. \ref{FIGspinvorticity} and see the contributions of $ \tilde{\zeta}^{f,\perp}_{s}$ and $\tilde{\xi}_s^{\omega,\perp}$ connecting vorticity and $\hat{\sv}$.
The driving field $\Ev$ includes all the forces acting on the electrons, such as friction force from the boundary, besides the applied external electric field. 
The field $\Ev$ is therefore locally proportional to the local fluid current $\jv$ neglecting the higher order contributions of the spin-orbit interaction, and is suppressed near the boundary as in Fig. \ref{FIGspinvorticity} \cite{FunakiAH21}.
This suppression  of fluid velocity in the $z$-direction leads to a vorticity along the $y$-direction, $\omega_y=\partial_z E_x$.
The vorticity changes sign on the upper and lower boundaries and changes along the $z$-direction. 
For electron spin along $+y$ direction, the conservative spin motive force is along the $z$ axis as 
$f^\zeta_z= \tilde{\zeta}^{f,\perp}_{s}\nabla_z \omegavE_y$, i.e., is in the $-z$ direction (if $\tilde{\zeta}^{f,\perp}_{s}>0$),
 while it is $+z$ direction for the electron spin pointing $-y$ direction.
Thus spin current along the $z$ axis is induced by the spin-vorticity coupling as argued in previous works \cite{MatsuoHydro17}.
The nonconservative motive force described by the term $\tilde{\xi}_s^{\omega,\perp}$, which is mainly from the asymmetric viscosity, acts in the $-y$ ($+y$) direction for spin polarization along $+z$ ($-z$) direction, inducing a spin current in the $y$ direction (Fig. \ref{FIGspinvorticity}). 
The contribution  $ \tilde{\zeta}^{f,\perp'}_{s}$ and $\tilde{\xi}_s^{\omega,\perp'}$ induce orthogonal spin polarizations to $ \tilde{\zeta}^{f,\perp}_{s}$ and $\tilde{\xi}_s^{\omega,\perp}$, respectively. 
The adiabatic contributions and anomalous Hall contributions are discussed by replasing $\hat{\sv}$ by $\Mvhat$ in the above argument. 

In the next section, we carry out microscopic calculation on a simplified model to confirm the above argument.

\subsection{Microscopic derivation \label{SECmicro}}
We first derive the definition of the spin-resolved momentum flux density by deriving a hydrodynamic equation for the time-derivative of momentum density, $\dot{\pv}$. 
In the present case with spin, spin-resolved momentum density, defined as 
\begin{align}
 p_i^\alpha\equiv \average{c^\dagger \hat{p}_i\sigma_\alpha c}, 
\end{align}
 is considered, where $\hat{p}_i$ and $\sigma_\alpha$ denote the operators for momentum and spin ($i$ and $\alpha$ represents direction), respectively, 
 $\average{\ }$ denotes quantum average and $c$ and $c^\dagger$ are electron field operators. 
Its time-derivative is derived by use of the Heisenberg equation of motion,
$\dot{p}_i^\alpha=i\average{[H,c^\dagger \hat{p}_i\sigma_\alpha c]}$, where $[A,B]\equiv AB-BA$ is a commutator and $H$ is the total Hamiltonian, as was done in the spinless cases \cite{Funaki21,FunakiAH21}. 
The Hamiltonian we consider is 
\begin{align}
 H &=
 \intr c^\dagger \lt( \frac{-\nabla^2}{2m}   -  (\Mv\cdot\sigmav) \rt) c +H_{\rm so}+H_{\rm i}
\end{align}
where $\Mv$ is a vector representing the magnetization including the exchange coupling constant.
The spin-orbit interaction by impurities is represented by 
\begin{align}
 H_{\rm so} 
 &= \lambda \int d^3r  c^\dagger(\rv) [(\nabla v(\rv) \times \hat{\pv}) \cdot\sigmav] c(\rv)  \nnr
&= i\lambda \sum_{\kv\kv'} v_{\kv'-\kv} (\kv'\times\kv) \cdot c^\dagger_{\kv'} \sigmav c_{\kv} 
\end{align}
where $\lambda$ is a coupling constant, $v(\rv)=v_{\rm i}\sum_{\Rv_n}\delta(\rv-\Rv_n)$ is an impurity potential, where $v_{\rm i}$ and $\Rv_n$ are the strength of the impurity potential and the position of  $n$-th impurity, respectively, and the impurity scattering potential is $H_{\rm i}= \int d^3r v(\rv)  c^\dagger(\rv) c(\rv)$.
The impurity scattering induces an electron lifetime of elastic scattering, $\tau$, given by
$\tau^{-1}=2\pi\nu\nimp\vi^2$, where $\nimp$ is the impurity concentration and $\nu$ is the density of states of electron. 

A driving electric  field for electron flow, $\Ev$, is included using a vector potential $\Av$ satisfying $\Ev=-\dot{\Av}$. 
As is known generally for transport theories, dominant nonequilibrium contributions for the case of time-independent $\Ev$ are those containing both retarded and advanced  Green's functions \cite{Funaki21}, and we shall focus on these contributions.  
Spatial inhomogeneity is taken into account by expanding response functions with respect to the wave vector $\qv$ of $\Ev$ to the linear order. 

In the present model, the driving field is not only the applied external field, but is an effective one that includes other extrinsic forces such as those introduced by boundaries. 
At the boundary, fluid velocity vanishes and this fact is imposed usually as a boundary condition in solving fluid dynamics.  
In the present microscopic modeling of the ohmic fluid, such boundary effects are taken account by assuming that the total driving field vanishes at the boundary as was done in Refs. \cite{Funaki21,FunakiAH21}. 
In fact, vanishing fluid velocity indicates that the force due to the applied field and the friction from the boundary cancel each other. 
In practice, our total effective field is related to the actual current density $\jv$ via a local relation $\Ev=(\sigmae)^{-1}\jv$, where $\sigmae$ is a conductivity tensor. 
In the present analysis focusing on the lowest order of the spin-orbit interaction, $\sigmae$ can be treated as diagonal.
It would be an interesting future work to solve numerically the whole hydrodynamic equation taking account of off-diagonal conductivity.

The hydrodynamic equation for spin-resolved momentum density is (see  Sec. \ref{SECpiderivation} for derivation) 
\begin{align}
 \dot{p}_i^\alpha &= - \nabla_j \pi_{ij}^{s,\alpha} +f_{s,i}^\alpha
 \label{sialphaeq}
\end{align}
where 
$ \pi^{s,\alpha}_{ij} \equiv  \pi^{0\alpha}_{ij}+ \delta\pi^{\alpha}_{ij} $
with 
\begin{align} 
 \pi^{0\alpha}_{ij}(\rv,t) &\equiv
 -i\tr\lt[\hat{p}_i\{ \hat{v}_j,\sigma_\alpha\} G^<(\rv,t,\rv,t) \rt]
 \label{pizerodef}
\end{align}
is a  conventional contribution in the form of momentum and velocity, while 
\begin{align} 
 \delta\pi^{\alpha}_{ij}(\rv,t) 
 &\equiv  -i\frac{\lambda}{2}\tr\lt[(\nabla_i\nabla_k V) (\delta_{j\alpha}\delta_{k\beta}-\delta_{j\beta}\delta_{k\alpha})
 \sigma_\beta G^<(\rv,t,\rv,t)\rt]
 \label{pideltadef}
\end{align}
is an anomalous contribution from the spin-orbit potential.
(tr is a trace over spin and $\{A,B\}\equiv AB+BA$.)
Here $G^<$ is the full lesser Green's function including the external field and 
\begin{align}
\hat{v}_j & =\frac{\hat{p}_j}{m}+\delta \hat{v}_j
\end{align}
is the total velocity operator with the anomalous velocity 
\begin{align}
\delta \hat{v}_j = -\lambda(\nabla V\times \sigmav)_j
 \end{align}
due to the spin-orbit interaction.
The term $f_{s,i}^\alpha$ in Eq. (\ref{sialphaeq}) is the one which cannot be written as a divergence of flow, interpreted purely as a force.
The spin-resolved momentum, which is essentially the spin current, is not conserved in the presence of spin relaxation processes, and this is why we have force density besides the momentum flux density.
This fact means that the definition of the spin-dependent momentum flux density is not unique. 
Nevertheless it is a useful quantity to study the roles of vorticity on the spin transport, like spin motive force (spin gauge field) in spintronics \cite{TataraReview19}. 

We calculate  the momentum flux density focus on the contribution from the spin-orbit interaction, as it is essential to couple electron flow and its spin. 
In the spin-orbit contribution of $\pi^{0\alpha}_{ij}$, there  are two processes, one arising from the normal velocity $\frac{\pv}{m}$, the contribution historically called skew scattering contribution, and the other arising from the anomalous velocity $\delta\vv$, called the side-jump contribution.   

The linear response contribution to the external field $\Ev$ is written as 
$\pi_{ij}^{s,\alpha}= \pi_{ijk}^{s,\alpha} E_k$, where $\pi_{ijk}^{s,\alpha}$ is a correlation function with one more current vertex for the applied field \cite{Funaki21}. 
Below, we evaluate dominant contributions to $\pi_{ijk}^{s,\alpha}$, which are those including both retarded and advanced Green's functions \cite{Funaki21}.

\subsubsection{Skew-scattering contribution}

\begin{figure}
 \includegraphics[width=0.2\hsize]{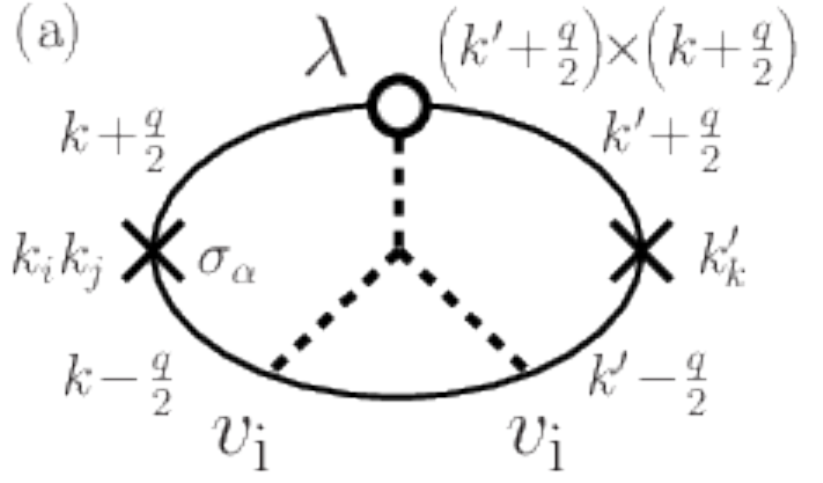}
 \includegraphics[width=0.2\hsize]{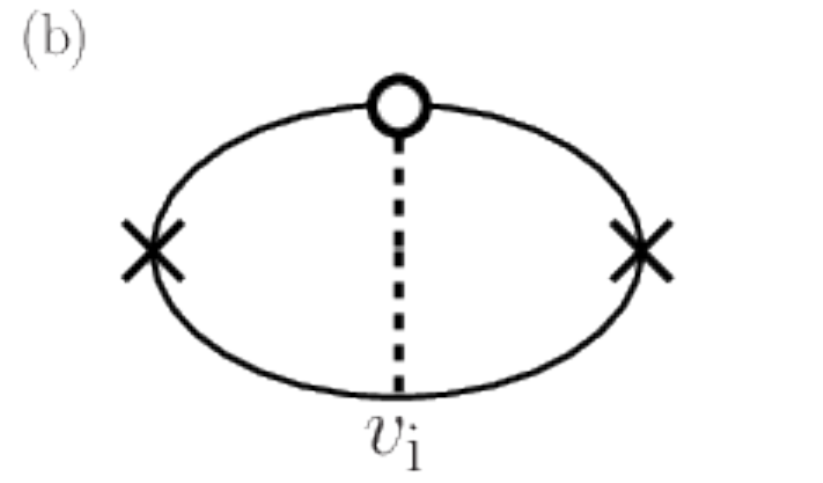}
 \includegraphics[width=0.2\hsize]{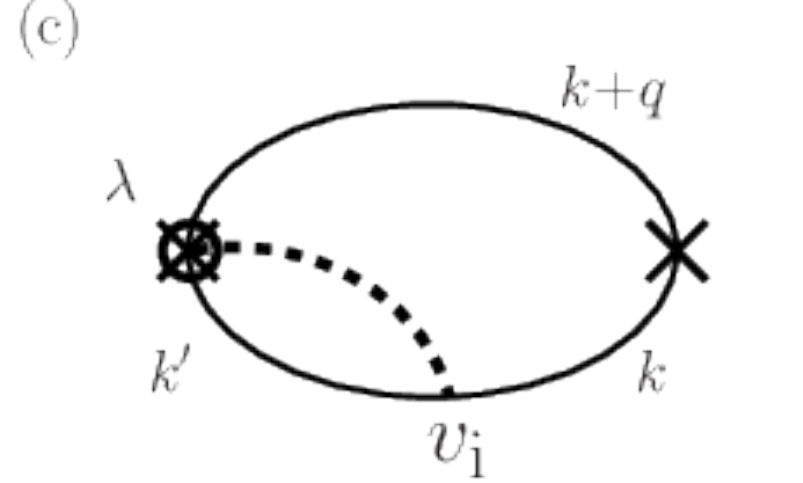}
 \includegraphics[width=0.2\hsize]{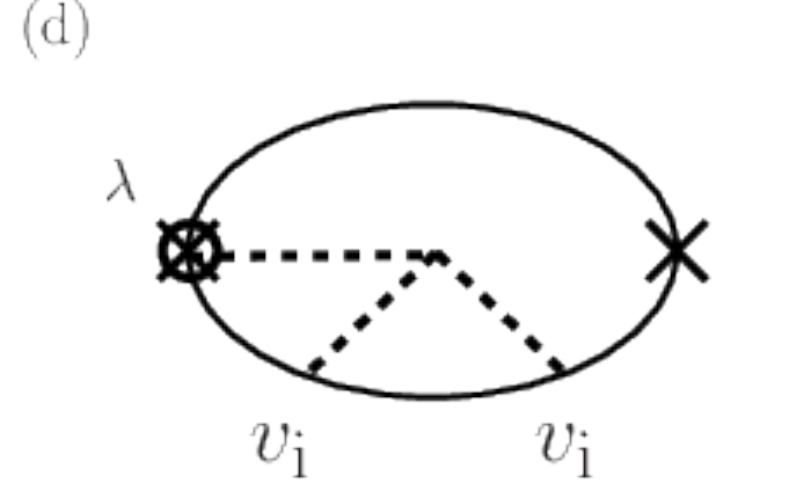}
 \caption{
Feynman diagrams for the dominant contribution to the spin-resolved momentum flux density $\pi^{s,\alpha}_{ijk}$  at the linear order in the spin-orbit interaction and linear response to the applied field, denoted by $\times$ at the right end (suffix $k$ represents the direction of the applied field). 
Solid lines represent free electron Green's functions, where upper and lower lines denote retarded ($g_{\kv}^\ret$) and advanced ($g_{\kv}^\adv$) Green's functions, respectively, and $\kv$ and $\kv'$ are the electron wave vectors. The wave vector $\qv$ is that of the external field and represents the inhomogeneity of  flow. 
Complex conjugate processes (turned upside down) are also taken into account.
(a): Skew-scattering contribution $\pi^{{\rm (ss)}\alpha}_{ijk}$.  The left vertex with $k_ik_j$ represents the vertex for the momentum flux density, and  a vertex with $\lambda$ denotes the spin-orbit interaction.
(b): A small contribution that is neglected.
(c): The contribution arising from the anomalous velocity $\delta v$ in $ \pi^{0\alpha}_{ijk}$, called the side-jump contribution. 
(d): The anomalous  contribution $\delta\pi^{\alpha}_{ijk} $. 
\label{FIGpiijdiag}}
\end{figure}

We first consider  the process of Fig.  \ref{FIGpiijdiag}(a) (skew scattering), which is 
\begin{align}
 \pi_{ijk}^{{\rm (ss)}\alpha}(\qv)
 &=\frac{e}{V^3}\frac{i}{\pi}\frac{\lambda \nimp\vi^3}{m^2} \Re \sum_{\kv\kv'}  
 k_i k_j k'_k\lt[\lt(\kv'+\frac{\qv}{2}\rt)\times\lt(\kv+\frac{\qv}{2}\rt)\rt]_\beta 
 \nnr
 & \times 
 \tr[\sigma_\alpha g_{\kv+\frac{\qv}{2}}^\ret \sigma_\beta g_{\kv'+\frac{\qv}{2}}^\ret
 g_{\kv'-\frac{\qv}{2}}^\adv g_{\kv''}^\adv g_{\kv-\frac{\qv}{2}}^\adv]  
 \label{pidef2}
\end{align}
where $\qv$ is the wave vector of the external field, $g_{\kv}^\ret \equiv [-\frac{k^2}{2m}+\ef+\Mv\cdot\sigmav+\frac{i}{2\tau}]^{-1}$ is  the free retarded green's function with elastic lifetime $\tau$ arising from the impurities and $g_{\kv}^\adv \equiv (g_{\kv}^\ret)^* $, $\ef$ being the Fermi energy.  
The real (imaginary) part is denoted by Re (Im).

We calculate the response function to the lowest order in the external wave vector $\qv$ and neglecting contributions smaller by a factor of $(\ef\tau)^{-1}$.
Noting that 
\begin{align}
\tr[\sigma^\alpha A \sigma^\beta B]  
&=
(\delta_{\alpha\beta} -\delta_{\alpha z} \delta_{\beta z})
\sum_{\sigma} A_\sigma B_{-\sigma}
+\delta_{\alpha z} \delta_{\beta z}
\sum_{\sigma} A_\sigma B_{\sigma} 
- i\epsilon_{\alpha\beta z} \sum_{\sigma} \sigma A_\sigma B_{-\sigma} 
\label{traceformula}
\end{align}
for diagonal matrices $A=\lt(\begin{array}{cc} A_+ & 0 \\ 0 & A_- \end{array}\rt)$ and $B$, the trace over the spin is calculated to obtain  (neglecting $O(q^2)$)
\begin{align}
 \pi_{ijk}^{{\rm (ss)}\alpha}(\qv)
&= i\sum_{\sigma}\Re[I_{ijk\alpha}^{-\sigma,\sigma} +\delta_{\alpha z} (I_{ijkz}^{\sigma,\sigma} -I_{ijkz}^{-\sigma,\sigma} ) +i\epsilon_{\alpha \beta z} \sigma I_{ijk\beta}^{-\sigma,\sigma} ]
\label{pidef22}
\end{align}
where 
$
I_{ijk\alpha}^{\sigma'\sigma}
\equiv \frac{1}{\pi}\frac{\lambda \nimp\vi^3}{m^2}\tilde{I}_{ijk\alpha}^{\sigma'\sigma}$, with 
\begin{align}
\tilde{I}_{ijk\alpha}^{\sigma'\sigma}
&= \frac{1}{V^3}\sum_{\kv\kv'\kv''}  
 k_i k_j k'_k(\kv'\times\kv)_\alpha  
 g_{\kv+\frac{\qv}{2},\sigma'}^\ret g_{\kv'+\frac{\qv}{2},\sigma}^\ret
 g_{\kv'-\frac{\qv}{2},\sigma}^\adv g_{\kv'',\sigma}^\adv g_{\kv-\frac{\qv}{2},\sigma}^\adv 
\end{align}
The summation over the wave vectors are carried out using 
\begin{align}
 \frac{1}{V} \sum_{\kv}g_{\kv,\sigma}^\adv &=i\pi\nu_\sigma, \,\,\,
 \frac{1}{V}\sum_{\kv}k_ik_jg_{\kv,\sigma}^\ret g_{\kv,\sigma}^\adv = \frac{2\pi}{3}\nu_\sigma k_{F\sigma}^2\tau \delta_{ij}
\nnr
\frac{1}{V} \sum_{\kv}k_ik_jk_k g_{\kv+\frac{\qv}{2},\sigma}^\ret g_{\kv-\frac{\qv}{2},\sigma'}^\adv
 &= 
 i\frac{\pi}{15m}(\delta_{ij}q_k+\delta_{ik}q_j+\delta_{jk}q_i) 
  \frac{{\nu_\sigma}k_{F\sigma}^4+\nu_{\sigma'}k_{F\sigma'}^4}{((\sigma-\sigma')M+\frac{i}{\tau})^2} 
  \nnr
\end{align}
where $\nu_\sigma$, $k_{F\sigma}$ are density of states at the Fermi energy and Fermi wave vector of electron with spin $\sigma=\pm$, respectively, as 
\begin{align}
\tilde{I}_{ijk\alpha}^{\sigma'\sigma}
 &= -\frac{2\pi^3}{45m} (\nu_{\sigma})^2 (k_{F\sigma})^2 \tau 
   \epsilon_{\alpha km}(\delta_{ij}q_m+\delta_{im}q_j+\delta_{jm}q_i) 
  \frac{{\nu_{\sigma'}}k_{F\sigma'}^4+\nu_{\sigma}k_{F\sigma}^4}{((\sigma'-\sigma)M+\frac{i}{\tau})^2} 
  \label{SSdominantcal2}
\end{align}
Thus 
\begin{align}
 \pi_{ijk}^{{\rm (ss)}\alpha}(\qv)
&= i\biggl[ [(\hat{\alphav}\times\qv)_k \delta_{ij}+\epsilon_{ik\alpha}q_j+\epsilon_{jk\alpha}q_i] \Re\sum_{\sigma}J_-^\sigma \nnr
&
+ \delta_{\alpha z}[(\hat{\zv}\times\qv)_k \delta_{ij}+\epsilon_{ikz}q_j+\epsilon_{jkz}q_i] \Re\sum_{\sigma}(J_+^\sigma-J_-^\sigma)
\nnr
& - \epsilon_{\alpha \beta z} [-\epsilon_{\beta kl} q_l \delta_{ij}+\epsilon_{ik\beta}q_j+\epsilon_{jk\beta}q_i] \Im\sum_{\sigma}\sigma J_-^\sigma\biggr]
\label{pidef222}
\end{align}
where 
\begin{align}
  J_-^\sigma
 &= \frac{2\pi^2}{45m^3}\lambda \nimp\vi^3 \lt(\sum_{\sigma'}{\nu_{\sigma'}}k_{F\sigma'}^4 \rt)
  (\nu_{\sigma})^2 (k_{F\sigma})^2 \tau 
  \frac{4M^2-\frac{1}{\tau^2}+4i\sigma \frac{M}{\tau}}{(4M^2+\frac{1}{\tau^2})^2} 
  \nnr
  J_+^\sigma
 &= -\frac{2\pi^2}{45m^3}\lambda  \nimp\vi^3 (\nu_{\sigma})^3 (k_{F\sigma})^6 2\tau^3 
\end{align}
We thus find that the result is consistent with the phenomenological argument of Sec. \ref{SEC:phenomenological};
We have 
\begin{align}
 \pi_{ij}^{{\rm (ss)}\alpha}
 &=
 \eta_{\parallel}^{\rm (ss)} \hat{\Mv}_\alpha 
 \lt( (\hat{\Mv}\cdot\omegavE)\delta_{ij} +\nabla_i T_j + \nabla_j T_i \rt)
\nnr
&+
\eta_{\perp}^{\rm (ss)} \lt( (\hat{\alphav}_\perp \cdot\omegavE)\delta_{ij} 
  + \nabla_i T^{\alpha_\perp}_{j} + \nabla_j T^{\alpha_\perp}_{i} \rt)
   \nnr
&+   
 \eta_{\perp'}^{\rm (ss)} \lt( (\hat{\alphav}_{\perp'} \cdot\omegavE)\delta_{ij} 
 + \nabla_i T^{\alpha_{\perp'}}_{j} + \nabla_j T^{\alpha_{\perp'}}_{i} \rt)
\end{align}
where $\hat{\alphav}$ is the unit vector along the spin direction $\alpha$, 
$\hat{\alphav}_{\perp}\equiv\hat{\alphav}-\Mvhat(\Mvhat\cdot  \hat{\alphav})$, 
$\hat{\alphav}_{\perp'}\equiv\Mvhat\times \hat{\alphav}$, and 
\begin{align}
\eta_{\parallel}^{\rm (ss)} 
&\equiv -\Re \sum_{\sigma}  J_+^\sigma \nnr
\eta_{\perp}^{\rm (ss)} 
&\equiv -\Re \sum_{\sigma}  J_-^\sigma \nnr
\eta_{\perp'}^{\rm (ss)} 
&\equiv -\Im \sum_{\sigma}  \sigma J_-^\sigma  
\end{align}
In terms of spin polarization vector $\svhat$ (Eq. (\ref{svhatdef})), the amplitude of the flux density is 
\begin{align}
 \overline{\pi}_{ij}^{{\rm (ss)}}
 &=
 \eta_{\parallel}^{\rm (ss)} (\svhat\cdot\hat{\Mv}) 
 \lt( (\hat{\Mv}\cdot\omegavE)\delta_{ij} +\nabla_i T_{\Mvhat,j} + \nabla_j T_{\Mvhat,i} \rt)
\nnr
&+
\eta_{\perp}^{\rm (ss)} \lt( (\hat{\sv}_\perp \cdot\omegavE)\delta_{ij} 
  + \nabla_i T_{\sv_\perp,j} + \nabla_j T_{\sv_\perp,i} \rt)
   \nnr
&+   
 \eta_{\perp'}^{\rm (ss)} \lt( (\hat{\sv}_{\perp'} \cdot\omegavE)\delta_{ij} 
 + \nabla_i T_{\sv_{\perp'},j} + \nabla_j T_{\sv_{\perp'},i} \rt)
\end{align}

Let us estimate the magnitude of the coefficients.
Our analysis in the ohmic regime (dirty metal) assumes $\ef\tau\gg1$.
We first consider strong ferromagnet with $M\tau\gg1$.
We simplify $\nu_+\sim\nu_-\sim1/\ef$ and $k_{F\sigma}\sim\kf$ for order of magnitude estimate and neglect numerical factors.
We then have 
\begin{align}
 J_+^\sigma & \sim \frac{\epsilon_{\rm so}}{m}\tau^2 , &
 J_-^\sigma & \sim \frac{\epsilon_{\rm so}}{m}\frac{1}{M^2}\lt(1+i\sigma\frac{1}{M\tau}\rt) 
\end{align}
and thus 
\begin{align}
\eta_{\parallel}^{\rm (ss)} 
&\sim  \frac{\epsilon_{\rm so}}{m}\tau^2, 
&
\eta_{\perp}^{\rm (ss)} 
& \sim  \frac{\epsilon_{\rm so}}{m}\frac{1}{M^2}, 
&
\eta_{\perp'}^{\rm (ss)} 
&\sim \frac{\epsilon_{\rm so}}{m}\frac{1}{M^3\tau}
\label{orderofestimatess}
\end{align}
where $\epsilon_{\rm so}\equiv \lambda \vimp \kf^2$ is the energy scale of the spin-orbit interaction. 
Namely, the adiabatic component is dominant and $\eta_{\perp}^{\rm (ss)} $ contribution is dominant in the nonadiabatic contributions.

In the limit of $M=0$, $J_+^\sigma=J_-^\sigma\sim  \frac{\epsilon_{\rm so}}{m}\tau^2$ and we have
\begin{align}
\eta_{\parallel}^{\rm (ss)} 
&=
\eta_{\perp}^{\rm (ss)} 
 \sim  \frac{\epsilon_{\rm so}}{m}\tau^2, &
\eta_{\perp'}^{\rm (ss)} =0,
\label{orderskewMzero}
\end{align}
which is natural from the rotational symmetry when $M=0$. 

There is a similar process with less impurity scattering, shown in Fig.  \ref{FIGpiijdiag}(b). 
Compared to the contribution $ \pi_{ij}^{{\rm (ss)}\alpha }$, it is without a factor of $i\pi\nu \vimp$.  Due to the extra factor of $i$, the imaginary and real parts are replaced, resulting in a vanishing contribution for $M=0$ and a reduction factor of $(M\tau)^{-1}$ in the case of $M\tau\gg1$.    
Considering a factor of $\nu\vimp\sim \sqrt{\nu\tau}$,  the contribution vanishes ($M=0$) or smaller by a factor of $(M\tau)^{-1/2}$.
We thus neglect this process.

\subsubsection{Side-jump and other contributions}

The contributions arising from the anomalous velocity $\delta \vv$ (the side-jump contribution, depicted in Fig. \ref{FIGpiijdiag}(c))  is 
\begin{align}
 \pi_{ijk}^{{\rm (sj)}\alpha}(\qv)
 &=
 \frac{-i}{4 \pi }\frac{1}{V^2}
 \frac{\lambda \nimp \vimp^{2} }{2m}
\epsilon_{\alpha jl} \sum_{\kv\kv'}
 {(k'+k+q)_i} \lt(k+\frac{q}{2}\rt)_k(k'-k)_l 
  \Re \tr[ g_{\kv'}^\ret g_{\kv}^\ret g_{\kv+\qv}^\adv]
 \label{sjdef1} 
\end{align}
It turns out to be 
\begin{align}
 \pi_{ijk}^{{\rm (sj)}\alpha}(\qv)&=  i\eta^{\rm (sj)}_0
  \lt(\frac{2}{3} \epsilon_{\alpha ij} q_k +\delta_{ik}\epsilon_{\alpha jl}q_l +\epsilon_{\alpha jk}q_i \rt)
 \label{sj1}
\end{align}
where
\begin{align}
 \eta^{\rm (sj)}_0 \equiv 
 \frac{\pi}{15}\frac{\lambda\nimp \vimp^{2} }{m^2}
 \sum_\sigma(\nu_\sigma)^2 k_{{\rm F}\sigma}^4\tau^2.
\end{align}
Thus, neglecting the term that vanishes in the force, we obtain 
\begin{align}
\pi_{ij}^{{\rm (sj)}} 
 &=  \eta^{{\rm (sj)},T_{\hat{\sv}}} (\nabla_i T_{\svhat,j} +\nabla_j T_{\svhat,i} ) +\epsilon_{ijk} a_k^{{\rm (sj)}}
\label{sj2}
\end{align}
where
\begin{align}
\av^{{\rm (sj)}}
 &= \xi^{{\rm (sj)},{\rm v}} \hat{\sv} (\nabla\cdot\Ev) + \xi^{{\rm (sj)},T_{\svhat}}(\nabla\times \Tv_{\svhat})
\end{align}
with $ \eta^{{\rm (sj)},T_{\hat{\sv}}} = -\frac{1}{2} \eta^{\rm (sj)}_0$, 
$ \xi^{{\rm (sj)},{\rm v}} = \frac{2}{3} \eta^{\rm (sj)}_0$, 
$\xi^{{\rm (sj)},T_{\svhat}}= -\frac{1}{2} \eta^{\rm (sj)}_0$. 
A unique feature of the side-jump contribution is that it does not vanish for $M=0$, in contrast to the skew scattering contribution. 
In the same order of estimate as in Eq. (\ref{orderofestimatess}), we have  
\begin{align}
 \eta^{\rm (sj)}_0 \sim   \frac{\epsilon_{\rm so}}{m}\sqrt{\frac{\tau^3}{\ef}} 
\end{align} 
The nonadiabatic contribution with spin polarized perpendicular to $\Mvhat$ is therefore dominated by the side-jump contribution instead of skew-scattering contribution  
represented by $\eta_{\perp}^{\rm (ss)} $ in the disordered metal with $M\tau\gg1$.  

Finally, the anomalous contribution to the momentum flux density is (Fig. \ref{FIGpiijdiag}(d))
\begin{align} 
 \delta\pi^{\alpha}_{ij}(\qv) 
 = \frac{(-i)^2}{2V^3 }
\frac{\lambda\nimp\vimp^{3}}{m}
\sum_{\kv\kv'\kv''} 
 (k'-k)_i (k'-k)_l \lt(k'+\frac{q}{2}\rt)_k 
 (\delta_{j\alpha}\delta_{l\beta}-\delta_{j\beta}\delta_{l\alpha})   
 \tr[\sigma_\beta (g_{\kv}^\ret g_{\kv''}^\ret + g_{\kv}^\adv g_{\kv''}^\adv  ) g_{\kv'}^\ret g_{\kv'+\qv}^\adv]
\end{align}
After some calculation, we obtain 
\begin{align}
\delta \pi_{ijk}^{\alpha}(\qv)
&=
 i\delta\pi \lt(\frac{8}{3}(\delta_{iz}\delta_{j\alpha}-\delta_{i\alpha}\delta_{jz}) q_k
 +\delta_{ik}(q_z\delta_{j\alpha}-q_\alpha\delta_{jz})
 - (\delta_{jz}\delta_{k\alpha}-\delta_{j\alpha}\delta_{kz}) q_i
 \rt)
 \end{align}
where
\begin{align}
 \delta\pi \equiv
 - \frac{2\pi^3}{15m^2}\lambda
{\nimp\vimp^{3} } \tau^2 \sum_{\sigma} \sigma \nu_{\sigma}^3  k_{F\sigma}^4
 \end{align}
Neglecting unphysical contribution that vanishes in the force, 
\begin{align}
\overline{\delta\pi}_{ij}
 &=  \delta \eta^{}_{T_{\perp'}} (\nabla_i T_{\tilde{\sv},j} +\nabla_j T_{\tilde{\sv},i} ) +\epsilon_{ijk} \delta a_k^{\tilde{\sv}}
\label{sj2}
\end{align}
where $\delta \eta^{}_{T_{\perp'}} =  \delta\pi $ and 
\begin{align}
\delta \av^{\tilde{\sv}}
 &= \delta\xi^{{\rm v}} \tilde{{\sv}} (\nabla\cdot\Ev) + \delta \xi^{T_{\tilde{\sv}}}(\nabla\times \Tv_{\tilde{\sv}})
\end{align}
with $\delta\xi^{{\rm v}}= \frac{8}{3} \delta\pi $, 
$ \delta \xi^{T_{\tilde{\sv}}}=  \delta\pi$. 
The order of  magnitude for $M\tau\gg1$ is $\delta\pi\sim \frac{\epsilon_{\rm so}}{m}\frac{\tau}{\ef}$ and is small than the side-jump contribution  but is larger than the skew scattering  nonadiabatic contribution.

In the limit of $M=0$, the order of magnitudes of $\eta^{\rm (sj)}_0$ and $ \delta\pi $ are the same as in the $M\tau\gg1$ case.
Thus these contributions are smaller than the skew scattering contribution for $M=0$, Eq. (\ref{orderskewMzero}). 
The side-jump contribution being smaller than the skew scattering one is due to the fact that the contribution involves less Green's functions than the skew scattering, resulting in smaller order of $\ef\tau(\gg1)$ considering the fact that the  peak magnitude of the Green's function is $\sim\tau$. 
When $M\tau\gg1$, in contrast, side-jump contribution dominates over the skew scattering as for the perpendicular spin polarization, as the Green's functions for the spin-flip processes are suppressed to be $\sim M^{-1}$, while side-jump process contains a contribution without spin flip (Eq. \ref{sjdef1}).

Phenomenological results in Sec. \ref{SEC:phenomenological} are therefore confirmed by microscopic calculations in this subsection.

\section{Vorticity-induced spin \label{SECs}}

Spin-resolved force density represents a force acting on spin components, namely, spin motive force driving spin current.
Although it would be interesting to explore spin transport phenomena solving the hydrodynamic equation with spin, we leave it as a future work and  study the result of the spin current generation, namely, the spin accumulation induced by the vorticity-induced  spin motive forces.   

Like spin current, spin-resolved force density is not  a clear physical observable, as the force acting on electrons with a particular spin polarization is not detectable. 
In contrast, spin density which we are going to calculate here is a physical observable. 
In fact, spin Hall effect originally argued as a relation between the spin density and an applied electric field by  Dyakonov and Perel \cite{Dyakonov71} is free from the ambiguity of definition of spin current \cite{TataraSH18}.
Here we discuss spin density taking account of inhomogeneous electric field to support physical consequence of spin hydrodynamic equations. 

\subsection{Spin Hall effect in the viewpoint of spin-vorticity coupling}
As was pointed out \cite{TataraSH18}, an inhomogeneous electric field applied to a metal with spin-orbit interaction induces a spin density as 
in Eq. (\ref{sh}). 
This relation in fact is a representation of spin Hall effect written in terms of spin density instead of spin current. 
The spin accumulation given by Eq. (\ref{sh}) in fact represents the spin accumulation formed at edges as a result of spin current generated by the applied electric field. 
The expression corresponds to the clean system where electron diffusion is not relevant and in the disordered case, the expression is multiplied by a diffusion propagator \cite{TataraSH18} (see eq. (\ref{snonlocal})). 
It indicates that spin Hall effect is a consequence of spin-vorticity coupling. 
The spin generation by the spin Hall effect has been argued mostly in the absence of magnetization, to discuss the pure spin current. 
When a magnetization is present, spin polarization along the magnetization $\hat{\Mv}$  (the adiabatic component) and orthogonal components $\hat{\Mv}\times\omegavE$ arise. 
We can therefore write 
\begin{align} %
 \sv(\rv)
  &=
 \lambda_{\rm sh} \omegavE 
 + \lambda_\parallel  \hat{\Mv}(\hat{\Mv}\cdot \omegavE)
  + \lambda_\perp (\hat{\Mv}\times\omegavE)
  \label{sinudedbyvorticity}
\end{align}
with coefficients $\lambda_\parallel$ and $\lambda_\perp$. 
The term  $\lambda_\parallel$ represents the adiabatic component (along $\Mv$) of the spin polarization and  $\lambda_\perp$ represents the anomalous Hall effect for the vorticity.

\subsection{Miscroscopic calculation}

Here we calculate the coefficients  based on the model of spin-orbit interaction due to heavy impurities of Sec. \ref{SECmicro}.
The dominant contribution arises from the same diagram as in Fig. \ref{FIGpiijdiag}(a) with the left vertex replaced by a Pauli matrix. 
The spin density polarized along direction $\alpha$ induced by the electric field along the $k$-direction is 
\begin{align}
 s_{k}^{ \alpha}(\qv)
 &=\frac{e}{V^3}\frac{i}{\pi}\frac{\lambda\nimp\vi^3}{m} \Re \sum_{\kv\kv'}  
 k'_k\lt[\lt(\kv'+\frac{\qv}{2}\rt)\times\lt(\kv+\frac{\qv}{2}\rt)\rt]_\beta
 \tr[\sigma_\alpha g_{\kv+\frac{\qv}{2}}^\ret \sigma_\beta g_{\kv'+\frac{\qv}{2}}^\ret
 g_{\kv'-\frac{\qv}{2}}^\adv g_{\kv''}^\adv g_{\kv-\frac{\qv}{2}}^\adv]  
 \label{pidef2}
\end{align}
which leads using Eq. (\ref{traceformula}) to 
\begin{align}
 s_{k}^{ \alpha}(\qv)&= 
 s_{k}^{{\rm sh},\alpha}(\qv)+\delta_{\alpha,z}  s_{k}^{{\parallel}}(\qv)+\epsilon_{\alpha\beta z} s_{k}^{{\perp}}(\qv) \nnr
 s_{k}^{{\rm sh},\alpha}(\qv)&= 
 \frac{e}{V^3}\frac{i}{\pi}\frac{\lambda \nimp\vi^3}{m} \Re \sum_{\kv\kv'\kv''}  
 k'_k\lt[\kv'\times\kv\rt]_\alpha \sum_{\sigma=\pm} g_{\kv+\frac{\qv}{2},\sigma}^\ret g_{\kv'+\frac{\qv}{2},-\sigma}^\ret
 g_{\kv'-\frac{\qv}{2},-\sigma}^\adv g_{\kv'',-\sigma}^\adv g_{\kv-\frac{\qv}{2},-\sigma}^\adv  \nnr
 s_{k}^{{\parallel}}(\qv)&= 
 \frac{e}{V^3}\frac{i}{\pi}\frac{\lambda \nimp \vi^3}{m} \Re \sum_{\kv\kv'\kv''}  
 k'_k\lt[\kv'\times\kv\rt]_z \sum_{\sigma=\pm} (g_{\kv+\frac{\qv}{2},\sigma}^\ret-g_{\kv+\frac{\qv}{2},-\sigma}^\ret) g_{\kv'+\frac{\qv}{2},\sigma}^\ret
 g_{\kv'-\frac{\qv}{2},\sigma}^\adv g_{\kv'',\sigma}^\adv g_{\kv-\frac{\qv}{2},\sigma}^\adv   \nnr
 s_{k}^{\perp}(\qv)&= 
 \frac{e}{V^3}\frac{i}{\pi}\frac{\lambda \nimp \vi^3}{m} \Re (-i)\sum_{\kv\kv'\kv''}  
 k'_k\lt[\kv'\times\kv\rt]_\beta \sum_{\sigma=\pm}\sigma g_{\kv+\frac{\qv}{2},\sigma}^\ret g_{\kv'+\frac{\qv}{2},-\sigma}^\ret
 g_{\kv'-\frac{\qv}{2},-\sigma}^\adv g_{\kv'',-\sigma}^\adv g_{\kv-\frac{\qv}{2},-\sigma}^\adv  
 \label{sdef3}
\end{align}
After summation over the wave vectors, the coefficients in Eq. (\ref{sinudedbyvorticity})  are obtained as 
\begin{align} %
\lambda_{\rm sh}  &= 
 e\frac{\pi^2}{3}\frac{\lambda  \nimp\vi^3}{m^2} 
\tau\frac{4M^2-\frac{1}{\tau^2}}{\lt(4M^2+\frac{1}{\tau^2}\rt)^2} \sum_{\sigma\sigma'} \nu_{\sigma} k_{F\sigma}^2 \nu_{\sigma'}^2 k_{F\sigma'}^2
\nnr
\lambda_\parallel  &= 
 e\frac{\pi^2}{3}\frac{\lambda  \nimp\vi^3}{m^2}
(-2\tau)
 \sum_{\sigma}\lt[  \nu_{\sigma}^3 k_{F\sigma}^4 \tau^2 + \nu_{\sigma}^2 k_{F\sigma}^2 \sum_{\sigma'}\nu_{\sigma'} k_{F\sigma'}^2 
 \frac{\lt(4M^2-\frac{1}{\tau^2}\rt)}{\lt(4M^2+\frac{1}{\tau^2}\rt)^2} \rt]
\nnr
 \lambda_\perp &= 
  e\frac{\pi^2}{3}\frac{\lambda \nimp \vi^3}{m^2} 
\frac{-4M}{\lt(4M^2+\frac{1}{\tau^2}\rt)^2} \sum_{\sigma\sigma'} \nu_{\sigma} k_{F\sigma}^2 \nu_{\sigma'}^2 k_{F\sigma'}^2 \sigma'
 \end{align}
In the disordered case with $M\tau\gg1$, the order of magnitude of each term  is 
\begin{align} %
\lambda_{\rm sh} &\sim \frac{\epsilon_{\rm so}}{\kf^2}\frac{1}{M^2}, &
\lambda_\parallel   &\sim \frac{\epsilon_{\rm so}}{\kf^2}\tau^2, &
 \lambda_\perp  &\sim \frac{\epsilon_{\rm so}}{\kf^2}\frac{1}{M^3\tau}
 \label{lamshmag}
\end{align}
meaning that the adiabatic spin polarization is dominant, while the spin Hall effect dominates the perpendicular component. 
In the limit of $M=0$,  $\lambda_{\rm sh}$ representing the spin-vorticity coupling and spin Hall effect  remains finite, while $\lambda_\parallel$ and $\lambda_\perp$ vanish.

\subsection{Vorticity-induced torque}
In ferromagnets, the vorticity-induced spin, Eq. (\ref{sinudedbyvorticity}), generates a current-induced torque near surfaces given by
\begin{align}
 \Tv_{\omega} &=\lambda_{\rm sh}[\hat{\Mv}\times (\nabla\times\Ev)]+\lambda_\perp[\hat{\Mv}\times[\hat{\Mv}\times (\nabla\times\Ev)]]
 \label{Tomega}
\end{align}
In ferromagnets with $M\tau \gg1$ and $M/\ef=O(1)$,  the first term dominates as is seen from Eq. (\ref{lamshmag}). 
Interestingly the coefficient $\lambda_{\rm sh}$ does not depend on $\tau$ in this regime, resulting in a universal torque when written in terms of the applied field $\Ev$. 
Although current-induced torques on magnetic textures have been discussed intensively, effect of vorticity and inhomogeneous current density has not been focused on. 
It is of interest to confirm experimentally the correlation between the vorticity-induced torque and the spin Hall effect, both determined by the coefficient $\lambda_{\rm sh}$. 

Let us compare the vorticity-induced torque to the so-called $\beta$-torque induced by spin relaxation and magnetization structure, 
\begin{align}
\Tv_{\beta}=\beta a^3[\hat{\Mv}\times(\jv\cdot\nabla)\hat{\Mv}], 
\end{align}
 where $\beta$ is a small constant representing the rate of spin relaxation and $a$ is the lattice constant. 
Its magnitude in common ferromagnets is typically of the order of $\beta\sim (M\tau_{\rm sf})^{-1}$, where 
$\tau_{\rm sf}^{-1}\sim \epsilon_{\rm so}^2/\ef$ in the case of spin-orbit interaction  \cite{KTS06,TE08}. 
In terms of the field $E$, 
$\Tv_{\beta}=\frac{e}{\kf^2}\widetilde{\beta}[\hat{\Mv}\times(\Ev\cdot\nabla)\hat{\Mv}]$, where 
$\widetilde{\beta}\equiv \frac{\kf^2}{e}\beta a^3\sigma_{\rm e}\sim \widetilde{\epsilon_{\rm so}}^2 \ef\tau$, with $\widetilde{\epsilon_{\rm so}}\equiv \epsilon_{\rm so}/\ef$.
Thus the ratio of the torque and the field is $\Tv_{\beta}/E\sim \frac{e}{\kf^2\ell_M}\widetilde{\epsilon_{\rm so}}^2 \ef\tau$, while for the vorticity-induced torque it is  
$\Tv_{\omega}/E\sim \frac{e}{\kf^2 \ell_\omega}\widetilde{\epsilon_{\rm so}}$, where $\ell_M$ is the length scale of magnetization structure and $\ell_\omega$ is the length scale of surface vorticity in the case of ohmic fluid. 
Considering the fact that the $\beta$ torque is second order in the spin-orbit interaction, while the vorticity torque is linear, surface vorticity may have a crucial role in current-induced torques. 
Even stronger effect is expected when surface roughness is considered, as roughness would henhance generation of vorticity. 
Nucleation of magnetic domain wall and skyrmion by artificial notches \cite{Yu20} would be carried out by the vorticity-induced torque as was argued in Ref. \cite{Fujimoto21}.

\begin{figure}
 \includegraphics[width=0.4\hsize]{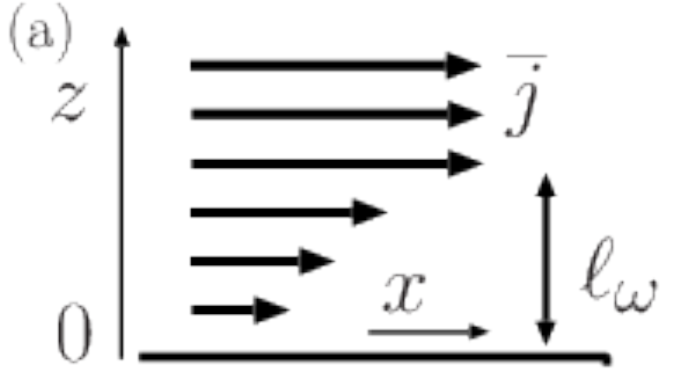}
 \includegraphics[width=0.4\hsize]{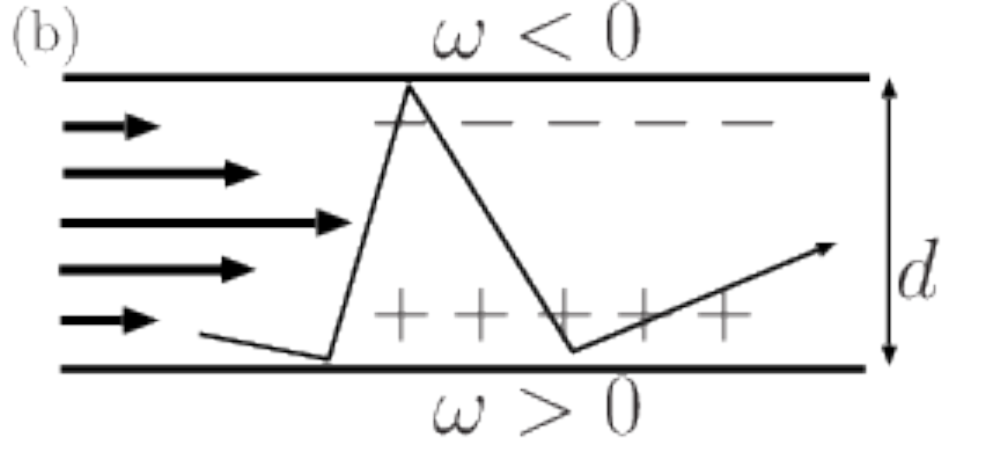}
 \caption{(a) Schematic figure showing the current profile near the boundary (chosen as $z=0$). Vorticity is finite in the length scale of $\ell_\omega$, where the current density changes. 
 (b)  Schematic figure showing vorticity-induced spin relaxation mechanism due to electron diffusion. The sign denotes the direction of vorticity $\omegav$. 
 Electron diffusion is represented by a zigzag line.
\label{FIGlomega}}
\end{figure}
Taking account of diffusion of electron, the spin density of Eq. (\ref{sinudedbyvorticity}) is modified to be long-ranged as \cite{TataraSH18}
\begin{align} %
 \sv(\rv)
  &= \int d^3r' D_s(\rv-\rv')[
 \lambda_{\rm sh} \omegavE(\rv') 
 + \lambda_\parallel  \hat{\Mv}(\hat{\Mv}\cdot \omegavE(\rv'))
  + \lambda_\perp (\hat{\Mv}\times\omegavE(\rv')) ]
  \label{snonlocal}
\end{align}
where $D_s(\rv)\equiv \frac{1}{\tau}\sum_{\qv}\frac{e^{i\qv\cdot\rv}}{Dq^2+\frac{1}{\tau_s}}$ is a diffusion propagator of electron spin, $D$ and $\tau_s$ are the diffusion constant and  spin relaxation time of electron, respectively.
The electron spin diffusion length is $\ell_s\equiv \sqrt{D\tau_s}$.
In the disordered case, therefore, the vorticity-induced torque is determined by thevorticity  average over the spin diffusion length, $\overline{\omegav}$.
Let us consider the case where $\ell_s$ is larger than the surface depth where the vorticity is finite, $\ell_\omega$. 
Choosing the axis as in Fig. \ref{FIGlomega}(a), local vorticity of current density $\jv$ is $\omega^{(j)}_y=\partial_z j_x$ and its average is 
\begin{align}
  \overline{\omega}^{(j)}_y=\frac{1}{\ell_\omega}\int_0^{\ell_\omega}dz \partial_z j_x =\frac{\overline{j}_x}{\ell_\omega}
\end{align}
where $\overline{j}_x$ is the current far from the surface and $j_x$ at the surface ($z=0$) is assumed to vanish.
The averaged torque near the surface,
\begin{align}
 \overline{\Tv}_{\omega} &=\lambda_{\rm sh}[\hat{\Mv}\times \overline{\omegav}]+\lambda_\perp[\hat{\Mv}\times[\hat{\Mv}\times \overline{\omegav}]]
 \label{Tomegaav}
\end{align}
is therefore 
\begin{align}
 \overline{\Tv}_{\omega} &= \frac{1}{ \sigmae} \lt[ \frac{\lambda_{\rm sh}}{\ell_\omega} [\hat{\Mv}\times (\zvhat\times\overline{\jv})]+\lambda_\perp[\hat{\Mv}\times[\hat{\Mv}\times  (\zvhat\times\overline{\jv})]] \rt]
 \label{Tomegares}
\end{align}
where $\zvhat$ denotes the direction normal to the surface. 
The torque represented by $\lambda_{\rm sh}$ has the same form as the one due to the surface Rashba-Edelstein effect, which induces spin polarization proportional to $\zvhat\times\jv$ \cite{Edelstein90}.

For thin film like in Fig. \ref{FIGspinvorticity}, the vorticity-induced torque points opposite on the upper and lower plane and has no bulk effects.

\subsection{Vorticity-induced spin relaxation}

The vorticity-induced torque is naturally inhomogeneous, and would lead to spin relaxation effects. 
Let us consider a thin film (Figs. \ref{FIGspinvorticity},\ref{FIGlomega}(b)) of thickness $d$ with electric field parallel to the magnetization. The vorticity is then  perpendicular to $\Mv$. 
The vorticity-induced torque drives the electron spin as 
\begin{align}
 \dot{\sv}=\gamma\lt[ 
 \lambda_{\rm sh}({\sv}\times \omegavE)+\lambda_{\perp}({\sv}\times\Mvhat)(\hat{\Mv}\cdot \omegavE)
 +\lambda_{\perp}[{\sv}\times(\hat{\Mv}\times \omegavE)] \rt], 
\end{align}
where $\gamma\equiv \frac{e}{m}$ is the electron gyromagnetoratio 
and 
thus the electron spin polarized along $\Mv$ gets flipped in a timescale of $(\lambda_{\rm sh}\omegaE)^{-1}$ (neglecting $\lambda_{\perp}(\ll \lambda_{\rm sh})$). 
In a thin film, vorticity $\omegavE$ arising from inhomogeneous flow points opposite on the upper and lower surfaces (Fig. \ref{FIGlomega}(b)). 
If call the length scale where vorticity is finite measured from the surface as $\ell_\omega$, 
(in the ohmnic fluid, $\ell_\omega\simeq \ell$ (mean free path)),  
$2\ell_\omega/d$ of the total electron spins are disturbed by the vorticity-induced torque.
The vorticity-induced spin relaxation time is therefore estimated as
\begin{align}
 {\tau_\omega}^{-1}
  &= \frac{\ell_\omega}{d} \lambda_{\rm sh} \omegaE
\end{align}
assuming that the spin diffusion length is longer than $d$. 
Its magnitude is as 
$ {\tau_\omega}^{-1} \sim  \widetilde{\epsilon_{\rm so}}\frac{eE}{\kf}\frac{1}{\kf d}$
 assuming $\omegaE\sim E/\ell_\omega$. 
For $E=10^{-2}$ V/$\mu$m and $\kf^{-1}\sim a=10^{-10}$m, 
$\frac{eE}{\kf}=10^{-6}$ eV=$2\times 10^{8}$ Hz. 
Let us define a damping constant as $\alpha_\omega\equiv {\tau_\omega}^{-1} t_M$, where $t_M$ is the time-scale of magnetization dynamics. 
For ${t_M}^{-1}\sim 1$ GHz, the above estimate leads to $\alpha_\omega\sim  0.2\times \frac{\widetilde{\epsilon_{\rm so}}}{\kf d}$. 
For a large spin-orbit metal of a thin film, in particular  with surface roughness, the vorticity-induced relaxation  may dominate over the intrinsic Gilbert damping. 
Experimentally, vorticity-induced damping would be separable from other isotropic intrinsic origins by using the anisotropic nature, i.e., vorticity-induced damping is suppressed if $\Ev \perp \Mv$.

\section{Vorticity-induced inverse spin Hall effect}
Spin current generation by fluid vorticity in a pipe was discussed in Ref. \cite{Matsuo17}, and the inverse spin Hall voltage due to the spin current and measured on heavy metal leads  was argued. 
In the case of the Hagen-Poiseuille flow considered there, the generated spin current is in the radial direction with spin polarization perpendicular to both radian and pipe directions and the inverse spin Hall voltage is along the flow.

The vorticity-induced inverse spin Hall effect is studied in the present context by calculating the spin-neutral force density (motive force for electric charge) taking account of the spin-orbit interaction to the second order in the absence of $M$. 
The inverse spin Hall voltage here is an intrinsic one without leads of heavy metals, and does not apply to experimental situations with heavy metal leads. 

The spin Hall coefficient ${\lambda}_{\rm sh}$ is a coefficient of the correlation function of spin and charge current density  linear in the external wave vector \cite{TataraSH18}.
The inverse spin Hall effect is represented as 
\begin{align}
\jv={\lambda}_{\rm ish}(\nabla\times\Bv_{\omega}), \label{jish}
\end{align}
where ${\lambda}_{\rm ish}$ is a coefficient proportional to $\lambda_{\rm sh}$ and  $\Bv_{\omega}$ is an effective magnetic field that drives spin density \cite{TataraSH18}. 
In the vorticity-driven case, the effective field is  induced by the vorticity as 
\begin{align}
\Bv_{\omega}= \frac{\lambda_{\rm sh}}{\chi}\omegavE, 
\label{Bomega}
\end{align}
 where $\chi$ is spin susceptibility. 
In the ohmic regime, the relation (\ref{jish}) indicates that there is a motive force $\fv^{\rm a}$ acting  on electron charge  that is proportional to the rotation of vorticity, $\nabla\times \omegavE$. 
The inverse spin Hall motive force is therefore represented by an antisymmetric component of the momentum flux density for charge (spin neutral), 
$\pi^{\rm a}_{ij}= \xi^{\rm a} \epsilon_{ijk} a_k$ with a vector $\av$, as $\fv^{\rm a}=-\xi^{\rm a}(\nabla\times\av)$, resulting in an inverse spin Hall current 
$\jv\propto(\nabla\times\av)$. 
As has been argued \cite{Groot11} (See also Eq. (\ref{piijnormal})), an antisymmetric component arises from vorticity, $\av\propto \omegavE$.
We thus obtain the inverse spin Hall current $\jv\propto\nabla\times\omegavE$,  consistent with Eqs. (\ref{jish})(\ref{Bomega}).

\begin{figure}
 \includegraphics[width=0.5\hsize]{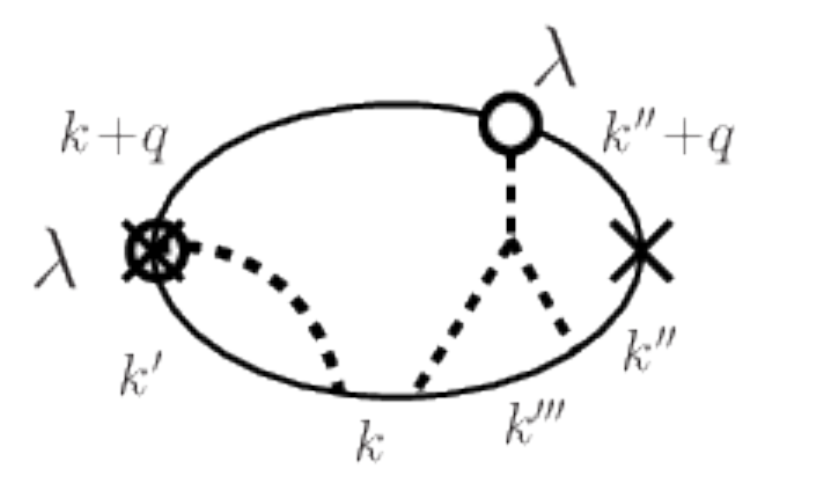}
 \caption{Feynman diagram representing the vorticity-induced inverse spin Hall effect at the second order of the spin-orbit interaction ($\lambda$). 
 The anomalous velocity vertex gives rise to an antisymmetric component of the momentum flux density that describes inverse spin Hall current, Eq. (\ref{jish}).
\label{FIGish}}
\end{figure}
Let us confirm this fact by a microscopic calculation of spin-neutral momentum flux density $\pi_{ij} \equiv \average{\hat{p}_i \hat{v}_j}\equiv \pi_{ijk}E_k$ in the case of $M=0$ including the spin-orbit interaction to the second order.
Obviously, the skew scattering contribution is symmetric with respect to $i$ and $j$ \cite{FunakiAH21}, and antisymmetic contribution arises from the side-jump contribution, whose dominant contribution is (shown in Fig. \ref{FIGish}) ($^{{\rm (sj)}(2)}$ denotes the side-jump process at the  second order of the spin-orbit interaction) 
\begin{align}
 \pi_{ijk}^{{\rm (sj)}(2)}(\qv)
 &=\frac{1}{4 \pi V^4}
 \frac{\lambda^2 \vimp^{5} n_{\rm i}^2}{2m}
i \epsilon_{\alpha jl}\epsilon_{mn\beta} 
\nnr & \times 
\sum_{\kv\kv'\kv''\kv'''}
 {(k'+k+q)_i} \lt(k''+\frac{q}{2}\rt)_k(k'-k)_l  k_n k''_m 
i \Im   \tr[ g_{\kv'}^\ret  \sigma_\alpha g_{\kv+\qv}^\adv \sigma_\beta g_{\kv''+\qv}^\adv g_{\kv''}^\ret  g_{\kv'''}^\ret g_{\kv}^\ret ]
 \label{sjso2def1} 
\end{align}
The antisymmetric component of $\pi_{ij}^{(2)}$  is written in terms of a vector $\av_{\omega}$ as 
$\pi_{ij}^{(2){\rm a}}=\epsilon_{ijk}a_{\omega,k}$, where 
\begin{align}
\av_{\omega}&= \xi^{(2)}_{\omega} \omegavE \nnr
 \xi^{(2)}_{\omega} &=
  \frac{\pi}{12} \lt(\frac{\epsilon_{\rm so}}{\eF}\rt)^2 D \tau^2 \vimp
=\lambda_{\rm sh} \lt(\frac{\epsilon_{\rm so}}{\eF}\rt)(\vimp\tau)\ef
\end{align}
The inverse spin Hall field induced by vorticity is therefore 
\begin{align}
\Ev_{\omega} = -\nabla\times \av_{\omega}
= -\xi^{(2)}_{\omega} \nabla\times \omegavE
\end{align}
The inverse spin Hall voltage due to the vorticity-induced spin current is therefore along the applied field $\Ev$ \cite{Matsuo11}. 
We note that the inverse spin Hall voltage in reality would be mixed with contributions from the symmetric components of the momentum flux density (see argument in Sec. \ref{SECah}). 

\section{Summary and discussion}
We have explored the effects of vorticity of electron flow in metals  on the spin transport based on a hydrodynamic viewpoint. 
The spin-resolved momentum flux density was discussed extending the results for anomalous Hall fluid studied previously in Ref. \cite{FunakiAH21}.
When represented by an external electric field $\Ev$, the anomalous Hall contributions to the momentum flux density are written in terms of a troidal moment, 
$\Tv_{\Mv}=\Mv\times\Ev$.
When electron spin is taken into account, the spin-resolved momentum flux density $\pi_{ij}^{\alpha}$ ($i$ and $j$ represent the spatial direction and $\alpha$ denotes spin direction) are characterized by the three contributions, namely, the adiabatic contribution where spin $\sv$ is parallel to $\Mv$, the nonadiabatic contribution written by $\sv_\perp\equiv \sv-\Mvhat(\Mvhat\cdot\sv)$  and a contribution written by $\tilde{\sv}\equiv \sv\times \Mv$. 
The troidal moment $\Tv_{\Mv}$ is  therefore extended to include two nonadiabatic contributions, $\Tv_{\sv_\perp}\equiv\sv_\perp\times\Ev$ and $\Tv_{\tilde{\sv}}\equiv\tilde{\sv}\times\Ev$.
Those results were confirmed by linear response calculations in a model with the impurity-induced spin-orbit interaction.

The spin-resolved force density $-\nabla_j\pi_{ij}^{\alpha}$, or spin motive force, was argued and roles of electron vorticity on the motive force was discussed.
Besides the conventional conservative force due to spin-vorticity coupling, proportional to $\nabla(\sv\cdot\omegav)$ ($\omegav$ is the vorticity), there is a nonconservative force, proportional to $\nabla\times(\omegav\times\sv)=(\sv\cdot\nabla)\omegav$.  
Like spin current, the spin-resolved momentum flux density and force density are physical conserved current, and their definition are not unique. 

Spin density induced by vorticity was studied. 
It was argued that the spin-vorticity coupling represents the spin Hall effect. 
The vorticity-induced spin density was discussed to give rise to a damping localized near the surface and interface where vorticity is induced.
The torque arises from the homogeneity of the current (or total field), in contrast to conventional current-induced relaxation torques such as $\beta$ torque arising from inhomogeneous magnetization. 
The vorticity-induced torque is a linear effect of spin-orbit interaction, while the conventional relaxation torque is second-order, meaning that the vorticity effects may dominate in thin films. 
The effect leads to torque and relaxation localized near  surfaces and interfaces. 
The inverse spin Hall effect due to the vorticity-induced spin current was shown to be described by the antisymmetric part of the momentum flux density evaluated at the second order of the spin-orbit interaction.

We considered the case of uniform magnetization with inhomogenuity of the applied electric field.
For a complete discussion of the troidal moment, inhomogeneous magnetization needs to be taken into account.
For this future work, an effective gauge field approach \cite{TataraReview19} is expected to be useful. 

\acknowledgements
The author thank  H. Funaki and R. Toshio for valuable discussion. 
This study was supported by
a Grant-in-Aid for Scientific Research (B) (No. 21H01034) from the Japan Society for the Promotion of Science.

\appendix
\section{Derivation of momentum flux density of Eqs. (\ref{pizerodef})
(\ref{pideltadef}) \label{SECpiderivation}}
Here we derive the expression for the spin-resolved momentum flux density by evaluating the time-derivative of the spin-resolved momentum, 
$\dot{p}_i^\alpha=i\average{[H,c^\dagger \hat{p}_i\sigma_\alpha c]}$.
The contribution from the spin-orbit coupling is focused. 
The commutator is calculated using $\nabla^{\rv}_i \delta(\rv-\rv')=-\nabla^{\rv'}_i \delta(\rv-\rv')$ and integral by parts
 as 
\begin{align}
  i&[H_{\rm so} ,(c^\dagger\hat{p}_i \sigma_\alpha c)_{\rv}] 
   =-\frac{i}{4}\lambda \int d\rv' \epsilon_{jkl}(\nabla^{\rv'}_j V(\rv')) \nnr
& \times
   \lt[ c^\dagger(\rv')\sigma_l (\nabla_k c(\rv'))-(\nabla_k c^\dagger(\rv'))\sigma_l c(\rv') ,
   c^\dagger(\rv)\sigma_\alpha (\nabla_i c(\rv))-(\nabla_i c^\dagger(\rv))\sigma_\alpha c(\rv) \rt]
   \nnr
    = & -\frac{i}{2}\lambda  
   \biggl[ (\nabla_j V)
   \biggl[ -\epsilon_{jk\alpha } \nabla_k [    c^\dagger \stackrel{\leftrightarrow}{\nabla_i} c ]
 + i \epsilon_{jkl}\epsilon_{l\alpha\beta} 
 [  c^\dagger \sigma_\beta \stackrel{\leftrightarrow}{\nabla_i} (\nabla_k c)  -(\nabla_k c^\dagger) \sigma_\beta \stackrel{\leftrightarrow}{\nabla_i} c ]  \biggr]
   \nnr
& + (\nabla_i \nabla_j V) \biggl[ -\epsilon_{jk\alpha } (c^\dagger  \stackrel{\leftrightarrow}{\nabla_k}  c )
 +i \epsilon_{jkl}\epsilon_{l\alpha\beta}   
 \nabla_k [ c^\dagger \sigma_\beta  c ] \biggr]
 \biggr]
 \nnr
 &  = -\nabla_j \hat{\pi}^{{\rm so},\alpha}_{ij} +\hat{f}^{{\rm so},\alpha}_i 
\end{align}
where 
\begin{align}
  \hat{\pi}^{{\rm so},\alpha}_{ij}  & =  \frac{i}{2}\lambda  
   \lt[ (\nabla_k V) \epsilon_{jk\alpha } [    c^\dagger \stackrel{\leftrightarrow}{\nabla_i} c ]
     - i(\nabla_i \nabla_k V) \epsilon_{jkl}\epsilon_{l\alpha\beta}    
  ( c^\dagger \sigma_\beta  c ) \rt]
   \label{pisodef}
   \\
 \hat{f}^{{\rm so},\alpha}_i &= \frac{i}{2}\lambda  
   \biggl[(\nabla_i \nabla_j V) \epsilon_{jk\alpha } (c^\dagger  \stackrel{\leftrightarrow}{\nabla_k}  c )
   -i (\nabla_j V) \epsilon_{jkl}\epsilon_{l\alpha\beta} 
 [  c^\dagger \sigma_\beta \stackrel{\leftrightarrow}{\nabla_i} (\nabla_k c)  -(\nabla_k c^\dagger) \sigma_\beta \stackrel{\leftrightarrow}{\nabla_i} c ]  
 \biggr] \label{fsodef}
\end{align}
are the spin-orbit contributions to the momentum flux density and force density, respectively.

The contribution of the kinetic term, $\hat{ \pi}^{K,\alpha}_{ij}$,  is similarly calculated, and the total momentum flux density operator is 
\begin{align}
 \hat{\pi}^{K,\alpha}_{ij}+\hat{\pi}^{{\rm so},\alpha}_{ij}(\rv,t) 
 &=   c^\dagger \biggl[ 
 \frac{1}{m} \hat{p}_i  \hat{p}_j \sigma_\alpha 
 -\lambda  
  (\nabla_k V) \epsilon_{jk\alpha }  \hat{p}_i 
  + \frac{1}{2}\lambda   (\nabla_i \nabla_k V) (\delta_{j\alpha}\delta_{k\beta}-\delta_{j\beta}\delta_{k\alpha})   
  \sigma_\beta   \biggr] c  \nnr
 & \equiv 
 \hat{\pi}^{s,\alpha}_{ij}
\end{align}
whose expectation value is  Eqs. (\ref{pizerodef})(\ref{pideltadef}).

The fact that the time-derivative of the spin-resolved momentum density is not written in terms of conserved current in the presence of spin relaxation, in the same way as the case of spin current. 
This means that the momentum flux density can not be defined uniquely. 
In other words, spin-resolved momentum density is not physical observable. 
In this paper, we use the form of Eq. (\ref{pisodef}). 
Different definitions result in different values of viscosity constants.
We note, however, that physical quantity like spin density are uniquely defined like in the spin current case \cite{TataraSH18}.


\begin{thebibliography}{26}%
\makeatletter
\providecommand \@ifxundefined [1]{%
 \@ifx{#1\undefined}
}%
\providecommand \@ifnum [1]{%
 \ifnum #1\expandafter \@firstoftwo
 \else \expandafter \@secondoftwo
 \fi
}%
\providecommand \@ifx [1]{%
 \ifx #1\expandafter \@firstoftwo
 \else \expandafter \@secondoftwo
 \fi
}%
\providecommand \natexlab [1]{#1}%
\providecommand \enquote  [1]{``#1''}%
\providecommand \bibnamefont  [1]{#1}%
\providecommand \bibfnamefont [1]{#1}%
\providecommand \citenamefont [1]{#1}%
\providecommand \href@noop [0]{\@secondoftwo}%
\providecommand \href [0]{\begingroup \@sanitize@url \@href}%
\providecommand \@href[1]{\@@startlink{#1}\@@href}%
\providecommand \@@href[1]{\endgroup#1\@@endlink}%
\providecommand \@sanitize@url [0]{\catcode `\\12\catcode `\$12\catcode
  `\&12\catcode `\#12\catcode `\^12\catcode `\_12\catcode `\%12\relax}%
\providecommand \@@startlink[1]{}%
\providecommand \@@endlink[0]{}%
\providecommand \url  [0]{\begingroup\@sanitize@url \@url }%
\providecommand \@url [1]{\endgroup\@href {#1}{\urlprefix }}%
\providecommand \urlprefix  [0]{URL }%
\providecommand \Eprint [0]{\href }%
\providecommand \doibase [0]{https://doi.org/}%
\providecommand \selectlanguage [0]{\@gobble}%
\providecommand \bibinfo  [0]{\@secondoftwo}%
\providecommand \bibfield  [0]{\@secondoftwo}%
\providecommand \translation [1]{[#1]}%
\providecommand \BibitemOpen [0]{}%
\providecommand \bibitemStop [0]{}%
\providecommand \bibitemNoStop [0]{.\EOS\space}%
\providecommand \EOS [0]{\spacefactor3000\relax}%
\providecommand \BibitemShut  [1]{\csname bibitem#1\endcsname}%
\let\auto@bib@innerbib\@empty
\bibitem [{\citenamefont {Baibich}\ \emph {et~al.}(1988)\citenamefont
  {Baibich}, \citenamefont {Broto}, \citenamefont {Fert}, \citenamefont
  {Van~Dau}, \citenamefont {Petroff}, \citenamefont {Etienne}, \citenamefont
  {Creuzet}, \citenamefont {Friederich},\ and\ \citenamefont
  {Chazelas}}]{Baibich88}%
  \BibitemOpen
  \bibfield  {author} {\bibinfo {author} {\bibfnamefont {M.~N.}\ \bibnamefont
  {Baibich}}, \bibinfo {author} {\bibfnamefont {J.~M.}\ \bibnamefont {Broto}},
  \bibinfo {author} {\bibfnamefont {A.}~\bibnamefont {Fert}}, \bibinfo {author}
  {\bibfnamefont {F.~N.}\ \bibnamefont {Van~Dau}}, \bibinfo {author}
  {\bibfnamefont {F.}~\bibnamefont {Petroff}}, \bibinfo {author} {\bibfnamefont
  {P.}~\bibnamefont {Etienne}}, \bibinfo {author} {\bibfnamefont
  {G.}~\bibnamefont {Creuzet}}, \bibinfo {author} {\bibfnamefont
  {A.}~\bibnamefont {Friederich}},\ and\ \bibinfo {author} {\bibfnamefont
  {J.}~\bibnamefont {Chazelas}},\ }\bibfield  {title} {\bibinfo {title} {Giant
  magnetoresistance of (001)fe/(001)cr magnetic superlattices},\ }\href
  {https://doi.org/10.1103/PhysRevLett.61.2472} {\bibfield  {journal} {\bibinfo
   {journal} {Phys. Rev. Lett.}\ }\textbf {\bibinfo {volume} {61}},\ \bibinfo
  {pages} {2472} (\bibinfo {year} {1988})}\BibitemShut {NoStop}%
\bibitem [{\citenamefont {Valet}\ and\ \citenamefont {Fert}(1993)}]{Valet93}%
  \BibitemOpen
  \bibfield  {author} {\bibinfo {author} {\bibfnamefont {T.}~\bibnamefont
  {Valet}}\ and\ \bibinfo {author} {\bibfnamefont {A.}~\bibnamefont {Fert}},\
  }\bibfield  {title} {\bibinfo {title} {Theory of the perpendicular
  magnetoresistance in magnetic multilayers},\ }\href
  {https://doi.org/10.1103/PhysRevB.48.7099} {\bibfield  {journal} {\bibinfo
  {journal} {Phys. Rev. B}\ }\textbf {\bibinfo {volume} {48}},\ \bibinfo
  {pages} {7099} (\bibinfo {year} {1993})}\BibitemShut {NoStop}%
\bibitem [{\citenamefont {Berger}(1978)}]{Berger78}%
  \BibitemOpen
  \bibfield  {author} {\bibinfo {author} {\bibfnamefont {L.}~\bibnamefont
  {Berger}},\ }\bibfield  {title} {\bibinfo {title} {Low‐field
  magnetoresistance and domain drag in ferromagnets},\ }\href
  {https://doi.org/10.1063/1.324716} {\bibfield  {journal} {\bibinfo  {journal}
  {Journal of Applied Physics}\ }\textbf {\bibinfo {volume} {49}},\ \bibinfo
  {pages} {2156} (\bibinfo {year} {1978})},\ \Eprint
  {https://arxiv.org/abs/https://doi.org/10.1063/1.324716}
  {https://doi.org/10.1063/1.324716} \BibitemShut {NoStop}%
\bibitem [{\citenamefont {Berger}(1986)}]{Berger86}%
  \BibitemOpen
  \bibfield  {author} {\bibinfo {author} {\bibfnamefont {L.}~\bibnamefont
  {Berger}},\ }\bibfield  {title} {\bibinfo {title} {Possible existence of a
  josephson effect in ferromagnets},\ }\href
  {http://link.aps.org/abstract/PRB/v33/p1572} {\bibfield  {journal} {\bibinfo
  {journal} {Phys. Rev. B}\ }\textbf {\bibinfo {volume} {33}},\ \bibinfo
  {pages} {1572} (\bibinfo {year} {1986})}\BibitemShut {NoStop}%
\bibitem [{\citenamefont {Tatara}\ \emph {et~al.}(2008)\citenamefont {Tatara},
  \citenamefont {Kohno},\ and\ \citenamefont {Shibata}}]{TKS_PR08}%
  \BibitemOpen
  \bibfield  {author} {\bibinfo {author} {\bibfnamefont {G.}~\bibnamefont
  {Tatara}}, \bibinfo {author} {\bibfnamefont {H.}~\bibnamefont {Kohno}},\ and\
  \bibinfo {author} {\bibfnamefont {J.}~\bibnamefont {Shibata}},\ }\bibfield
  {title} {\bibinfo {title} {Microscopic approach to current-driven domain wall
  dynamics},\ }\href {https://doi.org/doi:10.1016/j.physrep.2008.07.003}
  {\bibfield  {journal} {\bibinfo  {journal} {Physics Reports}\ }\textbf
  {\bibinfo {volume} {468}},\ \bibinfo {pages} {213} (\bibinfo {year}
  {2008})}\BibitemShut {NoStop}%
\bibitem [{\citenamefont {Matsuo}\ \emph {et~al.}(2011)\citenamefont {Matsuo},
  \citenamefont {Ieda}, \citenamefont {Saitoh},\ and\ \citenamefont
  {Maekawa}}]{Matsuo11}%
  \BibitemOpen
  \bibfield  {author} {\bibinfo {author} {\bibfnamefont {M.}~\bibnamefont
  {Matsuo}}, \bibinfo {author} {\bibfnamefont {J.}~\bibnamefont {Ieda}},
  \bibinfo {author} {\bibfnamefont {E.}~\bibnamefont {Saitoh}},\ and\ \bibinfo
  {author} {\bibfnamefont {S.}~\bibnamefont {Maekawa}},\ }\bibfield  {title}
  {\bibinfo {title} {Spin-dependent inertial force and spin current in
  accelerating systems},\ }\href {https://doi.org/10.1103/PhysRevB.84.104410}
  {\bibfield  {journal} {\bibinfo  {journal} {Phys. Rev. B}\ }\textbf {\bibinfo
  {volume} {84}},\ \bibinfo {pages} {104410} (\bibinfo {year}
  {2011})}\BibitemShut {NoStop}%
\bibitem [{\citenamefont {Matsuo}\ \emph
  {et~al.}(2017{\natexlab{a}})\citenamefont {Matsuo}, \citenamefont {Ohnuma},\
  and\ \citenamefont {Maekawa}}]{MatsuoHydro17}%
  \BibitemOpen
  \bibfield  {author} {\bibinfo {author} {\bibfnamefont {M.}~\bibnamefont
  {Matsuo}}, \bibinfo {author} {\bibfnamefont {Y.}~\bibnamefont {Ohnuma}},\
  and\ \bibinfo {author} {\bibfnamefont {S.}~\bibnamefont {Maekawa}},\
  }\bibfield  {title} {\bibinfo {title} {Theory of spin hydrodynamic
  generation},\ }\href {https://doi.org/10.1103/PhysRevB.96.020401} {\bibfield
  {journal} {\bibinfo  {journal} {Phys. Rev. B}\ }\textbf {\bibinfo {volume}
  {96}},\ \bibinfo {pages} {020401} (\bibinfo {year}
  {2017}{\natexlab{a}})}\BibitemShut {NoStop}%
\bibitem [{\citenamefont {Doornenbal}\ \emph {et~al.}(2019)\citenamefont
  {Doornenbal}, \citenamefont {Polini},\ and\ \citenamefont
  {Duine}}]{Doornenbal19}%
  \BibitemOpen
  \bibfield  {author} {\bibinfo {author} {\bibfnamefont {R.~J.}\ \bibnamefont
  {Doornenbal}}, \bibinfo {author} {\bibfnamefont {M.}~\bibnamefont {Polini}},\
  and\ \bibinfo {author} {\bibfnamefont {R.~A.}\ \bibnamefont {Duine}},\
  }\bibfield  {title} {\bibinfo {title} {Spin{\textendash}vorticity coupling in
  viscous electron fluids},\ }\href {https://doi.org/10.1088/2515-7639/aaf8fb}
  {\bibfield  {journal} {\bibinfo  {journal} {Journal of Physics: Materials}\
  }\textbf {\bibinfo {volume} {2}},\ \bibinfo {pages} {015006} (\bibinfo {year}
  {2019})}\BibitemShut {NoStop}%
\bibitem [{\citenamefont {Fujimoto}\ \emph {et~al.}(2021)\citenamefont
  {Fujimoto}, \citenamefont {Koshibae}, \citenamefont {Matsuo},\ and\
  \citenamefont {Maekawa}}]{Fujimoto21}%
  \BibitemOpen
  \bibfield  {author} {\bibinfo {author} {\bibfnamefont {J.}~\bibnamefont
  {Fujimoto}}, \bibinfo {author} {\bibfnamefont {W.}~\bibnamefont {Koshibae}},
  \bibinfo {author} {\bibfnamefont {M.}~\bibnamefont {Matsuo}},\ and\ \bibinfo
  {author} {\bibfnamefont {S.}~\bibnamefont {Maekawa}},\ }\bibfield  {title}
  {\bibinfo {title} {Zeeman coupling and dzyaloshinskii-moriya interaction
  driven by electric current vorticity},\ }\href
  {https://doi.org/10.1103/PhysRevB.103.L220402} {\bibfield  {journal}
  {\bibinfo  {journal} {Phys. Rev. B}\ }\textbf {\bibinfo {volume} {103}},\
  \bibinfo {pages} {L220402} (\bibinfo {year} {2021})}\BibitemShut {NoStop}%
\bibitem [{\citenamefont {Gurzhi}(1963)}]{Gurzhi63}%
  \BibitemOpen
  \bibfield  {author} {\bibinfo {author} {\bibfnamefont {R.}~\bibnamefont
  {Gurzhi}},\ }\bibfield  {title} {\bibinfo {title} {Minimum of resistance in
  impurity-free conductors},\ }\href
  {http://www.jetp.ac.ru/cgi-bin/e/index/e/17/2/p521?a=list} {\bibfield
  {journal} {\bibinfo  {journal} {JETP, Vol. 17, No. 2, p. 521 (August 1963)
  (Russian original - ZhETF, Vol. 44, No. 2, p. 771, August 1963 )}\ }
  (\bibinfo {year} {1963})}\BibitemShut {NoStop}%
\bibitem [{\citenamefont {Funaki}\ and\ \citenamefont
  {Tatara}(2021)}]{Funaki21}%
  \BibitemOpen
  \bibfield  {author} {\bibinfo {author} {\bibfnamefont {H.}~\bibnamefont
  {Funaki}}\ and\ \bibinfo {author} {\bibfnamefont {G.}~\bibnamefont
  {Tatara}},\ }\bibfield  {title} {\bibinfo {title} {Hydrodynamic theory of
  chiral angular momentum generation in metals},\ }\href
  {https://doi.org/10.1103/PhysRevResearch.3.023160} {\bibfield  {journal}
  {\bibinfo  {journal} {Phys. Rev. Research}\ }\textbf {\bibinfo {volume}
  {3}},\ \bibinfo {pages} {023160} (\bibinfo {year} {2021})}\BibitemShut
  {NoStop}%
\bibitem [{\citenamefont {Funaki}\ \emph {et~al.}(2021)\citenamefont {Funaki},
  \citenamefont {Toshio},\ and\ \citenamefont {Tatara}}]{FunakiAH21}%
  \BibitemOpen
  \bibfield  {author} {\bibinfo {author} {\bibfnamefont {H.}~\bibnamefont
  {Funaki}}, \bibinfo {author} {\bibfnamefont {R.}~\bibnamefont {Toshio}},\
  and\ \bibinfo {author} {\bibfnamefont {G.}~\bibnamefont {Tatara}},\
  }\bibfield  {title} {\bibinfo {title} {Vorticity-induced anomalous hall
  effect in an electron fluid},\ }\href
  {https://doi.org/10.1103/PhysRevResearch.3.033075} {\bibfield  {journal}
  {\bibinfo  {journal} {Phys. Rev. Research}\ }\textbf {\bibinfo {volume}
  {3}},\ \bibinfo {pages} {033075} (\bibinfo {year} {2021})}\BibitemShut
  {NoStop}%
\bibitem [{\citenamefont {Conti}\ and\ \citenamefont
  {Vignale}(1999)}]{Conti99}%
  \BibitemOpen
  \bibfield  {author} {\bibinfo {author} {\bibfnamefont {S.}~\bibnamefont
  {Conti}}\ and\ \bibinfo {author} {\bibfnamefont {G.}~\bibnamefont
  {Vignale}},\ }\bibfield  {title} {\bibinfo {title} {Elasticity of an electron
  liquid},\ }\href {https://doi.org/10.1103/PhysRevB.60.7966} {\bibfield
  {journal} {\bibinfo  {journal} {Phys. Rev. B}\ }\textbf {\bibinfo {volume}
  {60}},\ \bibinfo {pages} {7966} (\bibinfo {year} {1999})}\BibitemShut
  {NoStop}%
\bibitem [{\citenamefont {Tatara}(2018)}]{TataraSH18}%
  \BibitemOpen
  \bibfield  {author} {\bibinfo {author} {\bibfnamefont {G.}~\bibnamefont
  {Tatara}},\ }\bibfield  {title} {\bibinfo {title} {Spin correlation function
  theory of spin-charge conversion effects},\ }\href
  {https://doi.org/10.1103/PhysRevB.98.174422} {\bibfield  {journal} {\bibinfo
  {journal} {Phys. Rev. B}\ }\textbf {\bibinfo {volume} {98}},\ \bibinfo
  {pages} {174422} (\bibinfo {year} {2018})}\BibitemShut {NoStop}%
\bibitem [{\citenamefont {Dyakonov}\ and\ \citenamefont
  {Perel}(1971)}]{Dyakonov71}%
  \BibitemOpen
  \bibfield  {author} {\bibinfo {author} {\bibfnamefont {M.}~\bibnamefont
  {Dyakonov}}\ and\ \bibinfo {author} {\bibfnamefont {V.~I.}\ \bibnamefont
  {Perel}},\ }\bibfield  {title} {\bibinfo {title} {Possibility of orienting
  electron spins with current},\ }\href@noop {} {\bibfield  {journal} {\bibinfo
   {journal} {Sov. Phys. JETP Lett.}\ }\textbf {\bibinfo {volume} {13}},\
  \bibinfo {pages} {467} (\bibinfo {year} {1971})}\BibitemShut {NoStop}%
\bibitem [{\citenamefont {Hirsch}(1999)}]{Hirsch99}%
  \BibitemOpen
  \bibfield  {author} {\bibinfo {author} {\bibfnamefont {J.~E.}\ \bibnamefont
  {Hirsch}},\ }\bibfield  {title} {\bibinfo {title} {Spin hall effect},\ }\href
  {https://doi.org/10.1103/PhysRevLett.83.1834} {\bibfield  {journal} {\bibinfo
   {journal} {Phys. Rev. Lett.}\ }\textbf {\bibinfo {volume} {83}},\ \bibinfo
  {pages} {1834} (\bibinfo {year} {1999})}\BibitemShut {NoStop}%
\bibitem [{\citenamefont {Snider}\ and\ \citenamefont
  {Lewchuk}(1967)}]{Snider67}%
  \BibitemOpen
  \bibfield  {author} {\bibinfo {author} {\bibfnamefont {R.~F.}\ \bibnamefont
  {Snider}}\ and\ \bibinfo {author} {\bibfnamefont {K.~S.}\ \bibnamefont
  {Lewchuk}},\ }\bibfield  {title} {\bibinfo {title} {Irreversible
  thermodynamics of a fluid system with spin},\ }\href
  {https://doi.org/10.1063/1.1841187} {\bibfield  {journal} {\bibinfo
  {journal} {The Journal of Chemical Physics}\ }\textbf {\bibinfo {volume}
  {46}},\ \bibinfo {pages} {3163} (\bibinfo {year} {1967})},\ \Eprint
  {https://arxiv.org/abs/https://doi.org/10.1063/1.1841187}
  {https://doi.org/10.1063/1.1841187} \BibitemShut {NoStop}%
\bibitem [{\citenamefont {Scaffidi}\ \emph {et~al.}(2017)\citenamefont
  {Scaffidi}, \citenamefont {Nandi}, \citenamefont {Schmidt}, \citenamefont
  {Mackenzie},\ and\ \citenamefont {Moore}}]{Scaffidi17}%
  \BibitemOpen
  \bibfield  {author} {\bibinfo {author} {\bibfnamefont {T.}~\bibnamefont
  {Scaffidi}}, \bibinfo {author} {\bibfnamefont {N.}~\bibnamefont {Nandi}},
  \bibinfo {author} {\bibfnamefont {B.}~\bibnamefont {Schmidt}}, \bibinfo
  {author} {\bibfnamefont {A.~P.}\ \bibnamefont {Mackenzie}},\ and\ \bibinfo
  {author} {\bibfnamefont {J.~E.}\ \bibnamefont {Moore}},\ }\bibfield  {title}
  {\bibinfo {title} {Hydrodynamic electron flow and hall viscosity},\ }\href
  {https://doi.org/10.1103/PhysRevLett.118.226601} {\bibfield  {journal}
  {\bibinfo  {journal} {Phys. Rev. Lett.}\ }\textbf {\bibinfo {volume} {118}},\
  \bibinfo {pages} {226601} (\bibinfo {year} {2017})}\BibitemShut {NoStop}%
\bibitem [{\citenamefont {Landau}\ and\ \citenamefont
  {Lifshitz}(1987)}]{LandauLifshitz-FluidMechanics}%
  \BibitemOpen
  \bibfield  {author} {\bibinfo {author} {\bibfnamefont {L.~D.}\ \bibnamefont
  {Landau}}\ and\ \bibinfo {author} {\bibfnamefont {E.~M.}\ \bibnamefont
  {Lifshitz}},\ }\href@noop {} {\emph {\bibinfo {title} {Fluid Mechanics, 2nd
  ed.}}}\ (\bibinfo  {publisher} {(Butterworth-Heinemann, Oxford)},\ \bibinfo
  {year} {1987})\BibitemShut {NoStop}%
\bibitem [{\citenamefont {Groot}\ and\ \citenamefont {Mazur}(2011)}]{Groot11}%
  \BibitemOpen
  \bibfield  {author} {\bibinfo {author} {\bibfnamefont {S.~R.~D.}\
  \bibnamefont {Groot}}\ and\ \bibinfo {author} {\bibfnamefont
  {P.}~\bibnamefont {Mazur}},\ }\href@noop {} {\emph {\bibinfo {title}
  {Non-Equilibrium Thermodynamics}}}\ (\bibinfo  {publisher} {Dover Books on
  Physics},\ \bibinfo {year} {2011})\BibitemShut {NoStop}%
\bibitem [{\citenamefont {Tatara}(2019)}]{TataraReview19}%
  \BibitemOpen
  \bibfield  {author} {\bibinfo {author} {\bibfnamefont {G.}~\bibnamefont
  {Tatara}},\ }\bibfield  {title} {\bibinfo {title} {Effective gauge field
  theory of spintronics},\ }\href
  {https://doi.org/https://doi.org/10.1016/j.physe.2018.05.011} {\bibfield
  {journal} {\bibinfo  {journal} {Physica E: Low-dimensional Systems and
  Nanostructures}\ }\textbf {\bibinfo {volume} {106}},\ \bibinfo {pages} {208 }
  (\bibinfo {year} {2019})}\BibitemShut {NoStop}%
\bibitem [{\citenamefont {Kohno}\ \emph {et~al.}(2006)\citenamefont {Kohno},
  \citenamefont {Tatara},\ and\ \citenamefont {Shibata}}]{KTS06}%
  \BibitemOpen
  \bibfield  {author} {\bibinfo {author} {\bibfnamefont {H.}~\bibnamefont
  {Kohno}}, \bibinfo {author} {\bibfnamefont {G.}~\bibnamefont {Tatara}},\ and\
  \bibinfo {author} {\bibfnamefont {J.}~\bibnamefont {Shibata}},\ }\bibfield
  {title} {\bibinfo {title} {Microscopic calculation of spin torques in
  disordered ferromagnets},\ }\href {https://doi.org/10.1143/JPSJ.75.113706}
  {\bibfield  {journal} {\bibinfo  {journal} {Journal of the Physical Society
  of Japan}\ }\textbf {\bibinfo {volume} {75}},\ \bibinfo {pages} {113706}
  (\bibinfo {year} {2006})}\BibitemShut {NoStop}%
\bibitem [{\citenamefont {Tatara}\ and\ \citenamefont {Entel}(2008)}]{TE08}%
  \BibitemOpen
  \bibfield  {author} {\bibinfo {author} {\bibfnamefont {G.}~\bibnamefont
  {Tatara}}\ and\ \bibinfo {author} {\bibfnamefont {P.}~\bibnamefont {Entel}},\
  }\bibfield  {title} {\bibinfo {title} {Calculation of current-induced torque
  from spin continuity equation},\ }\href
  {https://doi.org/10.1103/PhysRevB.78.064429} {\bibfield  {journal} {\bibinfo
  {journal} {Phys. Rev. B}\ }\textbf {\bibinfo {volume} {78}},\ \bibinfo {eid}
  {064429} (\bibinfo {year} {2008})}\BibitemShut {NoStop}%
\bibitem [{\citenamefont {Yu}\ \emph {et~al.}(2020)\citenamefont {Yu},
  \citenamefont {Morikawa}, \citenamefont {Nakajima}, \citenamefont {Shibata},
  \citenamefont {Kanazawa}, \citenamefont {Arima}, \citenamefont {Nagaosa},\
  and\ \citenamefont {Tokura}}]{Yu20}%
  \BibitemOpen
  \bibfield  {author} {\bibinfo {author} {\bibfnamefont {X.~Z.}\ \bibnamefont
  {Yu}}, \bibinfo {author} {\bibfnamefont {D.}~\bibnamefont {Morikawa}},
  \bibinfo {author} {\bibfnamefont {K.}~\bibnamefont {Nakajima}}, \bibinfo
  {author} {\bibfnamefont {K.}~\bibnamefont {Shibata}}, \bibinfo {author}
  {\bibfnamefont {N.}~\bibnamefont {Kanazawa}}, \bibinfo {author}
  {\bibfnamefont {T.}~\bibnamefont {Arima}}, \bibinfo {author} {\bibfnamefont
  {N.}~\bibnamefont {Nagaosa}},\ and\ \bibinfo {author} {\bibfnamefont
  {Y.}~\bibnamefont {Tokura}},\ }\bibfield  {title} {\bibinfo {title} {Motion
  tracking of 80-nm-size skyrmions upon directional current injections},\
  }\bibfield  {journal} {\bibinfo  {journal} {Science Advances}\ }\textbf
  {\bibinfo {volume} {6}},\ \href {https://doi.org/10.1126/sciadv.aaz9744}
  {10.1126/sciadv.aaz9744} (\bibinfo {year} {2020}),\ \Eprint
  {https://arxiv.org/abs/https://advances.sciencemag.org/content/6/25/eaaz9744.full.pdf}
  {https://advances.sciencemag.org/content/6/25/eaaz9744.full.pdf} \BibitemShut
  {NoStop}%
\bibitem [{\citenamefont {Edelstein}(1990)}]{Edelstein90}%
  \BibitemOpen
  \bibfield  {author} {\bibinfo {author} {\bibfnamefont {V.}~\bibnamefont
  {Edelstein}},\ }\bibfield  {title} {\bibinfo {title} {Spin polarization of
  conduction electrons induced by electric current in two-dimensional
  asymmetric electron systems},\ }\href
  {https://doi.org/10.1016/0038-1098(90)90963-C} {\bibfield  {journal}
  {\bibinfo  {journal} {Solid State Communications}\ }\textbf {\bibinfo
  {volume} {73}},\ \bibinfo {pages} {233 } (\bibinfo {year}
  {1990})}\BibitemShut {NoStop}%
\bibitem [{\citenamefont {Matsuo}\ \emph
  {et~al.}(2017{\natexlab{b}})\citenamefont {Matsuo}, \citenamefont {Saitoh},\
  and\ \citenamefont {Maekawa}}]{Matsuo17}%
  \BibitemOpen
  \bibfield  {author} {\bibinfo {author} {\bibfnamefont {M.}~\bibnamefont
  {Matsuo}}, \bibinfo {author} {\bibfnamefont {E.}~\bibnamefont {Saitoh}},\
  and\ \bibinfo {author} {\bibfnamefont {S.}~\bibnamefont {Maekawa}},\
  }\bibfield  {title} {\bibinfo {title} {Spin-mechatronics},\ }\href
  {https://doi.org/10.7566/JPSJ.86.011011} {\bibfield  {journal} {\bibinfo
  {journal} {Journal of the Physical Society of Japan}\ }\textbf {\bibinfo
  {volume} {86}},\ \bibinfo {pages} {011011} (\bibinfo {year}
  {2017}{\natexlab{b}})},\ \Eprint
  {https://arxiv.org/abs/https://doi.org/10.7566/JPSJ.86.011011}
  {https://doi.org/10.7566/JPSJ.86.011011} \BibitemShut {NoStop}%
\end{thebibliography}
%

\end{document}